\documentclass[prb,twocolumn,superscriptaddress,citeautoscript,amsart]{revtex4-2}

\usepackage{graphicx}
\usepackage{multirow}
\usepackage{color}
\usepackage{bm}
\usepackage{times}
\usepackage{amsmath,bm,amsfonts}
\usepackage{dcolumn}
\usepackage{graphicx}
\usepackage{latexsym}
\usepackage{hhline}
\usepackage{braket}
\usepackage{bbold}
\usepackage{multirow}

\def\bb{\mathbb}

\begin{document}

\title{
Unconventional Majorana Fermions on the Surface of Topological Superconductors Protected by Rotational Symmetry
}

\author{Junyeong \surname{Ahn}}
\email{Present address: Department of Physics, Harvard University, Cambridge, Massachusetts 02138, USA\\
junyeongahn@fas.harvard.edu}
\affiliation{Center for Correlated Electron Systems, Institute for Basic Science (IBS), Seoul 08826, Korea}
\affiliation{Department of Physics and Astronomy, Seoul National University, Seoul 08826, Korea}
\affiliation{Center for Theoretical Physics (CTP), Seoul National University, Seoul 08826, Korea}
\affiliation{RIKEN Center for Emergent Matter Science (CEMS), Wako, Saitama 351-0198, Japan}
\affiliation{Department of Applied Physics, The University of Tokyo, Bunkyo, Tokyo 113-8656, Japan}

\author{Bohm-Jung\surname{Yang}}
\email{bjyang@snu.ac.kr}
\affiliation{Center for Correlated Electron Systems, Institute for Basic Science (IBS), Seoul 08826, Korea}
\affiliation{Department of Physics and Astronomy, Seoul National University, Seoul 08826, Korea}
\affiliation{Center for Theoretical Physics (CTP), Seoul National University, Seoul 08826, Korea}

\date{\today}

\begin{abstract}
Topological superconductors are exotic gapped phases of matter hosting Majorana mid-gap states on their boundaries. 
In conventional three-dimensional topological superconductors, Majorana in-gap states appear in the form of spin-1/2 fermions with a quasi-relativistic dispersion relation.
Here, we show that unconventional Majorana states can emerge on the surface of three-dimensional topological superconductors protected by rotational symmetry. 
The unconventional Majorana surface states are classified into three different categories: a spin-$s$ Majorana fermion with $(2s+1)$-fold degeneracy $(s\geq3/2)$, a Majorana Fermi line carrying two distinct topological charges, and a quartet of spin-1/2 Majorana fermions related by fourfold rotational symmetry. The spectral properties of the first two types, which go beyond conventional spin-1/2 fermions, are unique to topological superconductors and have no counterparts in topological insulators.
We show that unconventional Majorana surface states can be obtained in the superconducting phase of doped $Z_2$ topological insulators or Dirac semimetals with rotational symmetry. 
\end{abstract}


\maketitle

\section{Introduction}
 
Topologically stable gapless surface states are the hallmark of three-dimensional (3D) topological insulators (TIs) and topological superconductors (TSCs)~\cite{hasan2010colloquium}.
One common feature of such surface states is that they appear as spin-1/2 fermions with a quasi-relativistic dispersion relation.
According to the recent classification of TI surface states using wallpaper groups~\cite{wieder2018wallpaper}, gapless surface states of TIs always have the form of Dirac or Weyl fermions locally, while their global band structure can take various forms~\cite{wang2016hourglass,alexandradinata2016topological,wieder2018wallpaper}.
However, as crystalline systems do not have Lorentz symmetry, there is no fundamental reason forbidding more exotic dispersion relations.
Indeed, the discovery of exotic low-energy excitations in bulk semimetals and bulk nodes of superconductors, such as spin-1 and spin-3/2 fermions~\cite{kennett2011birefringent,bradlyn2016beyond,liang2016semimetal,tang2017multiple,hu2018topological}, nodal lines~\cite{zhang2013time,kim2015dirac,fang2015topological} and nodal surfaces~\cite{agterberg2017bogoliubov,bzdusek2017robust}, have shown that unconventional fermionic excitations protected by crystalline symmetries can emerge in bulk crystals.
Also, in contrast to TI surface states, symmetry-protected surface Majorana fermions (MFs) of TSCs have not yet been exhaustively characterized. 
Considering that TSCs have particle-hole symmetry that is absent in TIs, the surfaces of TSCs may host unusual fermions, which go beyond spin-1/2 fermions on TI surfaces.
 
\begin{figure}[t!]
\includegraphics[width=8.5 cm]{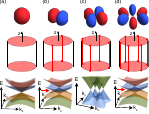}
\caption{
Majorana boundary states of rotation-protected topological superconductors in 3D.
The top panel shows the symmetry of the pairing function under $C_{n}$ rotation:
(a) even-$C_{n=2,3,4,6}$, and (b,c,d) odd-$C_{2,4,6}$.
The middle shows the real-space geometry of the system.
Here, the $z$-axis is the rotation axis.
Red regions host gapless Majorana fermions.
The characteristic surface spectra on $C_{n}$-invariant surfaces are shown at the bottom.
In (a), only the spectrum with a fourfold degenerate point is shown for clarity, but arbitrary degeneracy can be protected.
The red arrow in (b,d) indicates the zero energy where the MFL exists.
}
\label{fig:scheme}
\end{figure}

Here, we show that unconventional MFs emerge on the surfaces of TSCs protected by $n$-fold rotation $C_n$ and time-reversal $T$ symmetries.
By analyzing all possible realizations of anomalous surface states, we find that rotation-protected TSCs feature three types of surface MFs, two of which exhibit characteristic energy spectra that have no counterparts in TIs.
The first type takes the form of {\it higher-spin Majorana fermions} (HSMFs), which generalize the spin-3/2 fermion in semimetals~\cite{kennett2011birefringent,bradlyn2016beyond,liang2016semimetal,tang2017multiple} when the superconducting pairing function is invariant under $C_{n}$ (we call this even-$C_{n}$ pairing) [see Fig.~\ref{fig:scheme}(a)].
As higher-spin states cannot be realized on the boundaries of TIs with any wallpaper symmetry groups~\cite{wieder2018wallpaper}, they are unique to TSCs.
Furthermore, the HSMF cannot exist in the bulk of isolated two-dimensional (2D) nodal superconductors because their protection requires an anomalous $C_{n}$ symmetry representation.
On the other hand, when the pairing function changes sign under $C_{n=2,6}$ (we call this odd-$C_{n=2,6}$ pairing), a {\it doubly charged Majorana Fermi line} (DCMFL), carrying both zero-dimensional (0D) and one-dimensional (1D) topological charges, appears [Fig.~\ref{fig:scheme}(b,d)].
While the 0D topological charge indicates the local stability of the DCMFL, the 1D topological charge guarantees its global stability~\cite{bzdusek2017robust}.
Finally, when the pairing function changes its sign under $C_4$ (odd-$C_4$ pairing), a quartet of Majorana fermions (QMF) with twofold degeneracy appears on a $C_4$ invariant surface [Fig.~\ref{fig:scheme}(c)]
This is a superconducting analog of the $C_4$ rotation anomaly that was recently proposed in TIs protected by $C_n$ and $T$ symmetries~\cite{fang2019new}.
We show that all three types can appear when superconductivity emerges in doped ${\mathbb Z}_2$ TIs or Dirac semimetals with $T$ and $C_n$ symmetries.

To convey the main ideas concisely, we focus on spin-orbit coupled systems below. 
However, our theory is also applicable to spin-rotation-symmetric and spin-polarized systems, as explained in detail in the Supplemental Material (SM)~\cite{supp}. 
Our surface-state classification is consistent with previous bulk classifications~\cite{shiozaki2014topology,cornfeld2019classification,shiozaki2019classification,fang2017topological}, which shows that we have exhausted all possible anomalous surface states.

\section{Result}

{\it Formalism.---}
We consider the mean-field Hamiltonian for superconductors in the Bogoliubov-de Gennes (BdG) formalism,
$\hat{H}=\frac{1}{2}\sum_{\bf k}\hat{\Psi}^{\dagger}_{\bf k}H_{\rm BdG}({\bf k})\hat{\Psi}_{\bf k}$,
where
\begin{align}
H_{\rm BdG}({\bf k})
&=
\begin{pmatrix}
h({\bf k}) & \Delta({\bf k}) \\
\Delta^{\dagger}({\bf k})  & -\sigma_yh^t(-{\bf k})\sigma_y
\end{pmatrix},
\end{align}
and
$\hat{\Psi}
=(\hat{c}_{\bf k},\hat{c}^{\dagger}_{-{\bf k}}i\sigma_y)^t$ is the Nambu spinor in which $\hat{c}_{\bf k}/\hat{c}^{\dagger}_{\bf k}$ are electron annihilation/creation operators. The superscript $t$ denotes the matrix transpose. 
$h({\bf k})$ is the normal-state Hamiltonian, and the superconducting pairing function $\Delta({\bf k})$ satisfies $\Delta({\bf k})=-\sigma_y\Delta^t(-{\bf k})\sigma_y$ due to the Fermi statistics. 
This BdG Hamiltonian has particle-hole $P$ symmetry $PH_{\rm BdG}({\bf k})P^{-1}=-H_{\rm BdG}(-{\bf k})$ where
\begin{align}
P=
\begin{pmatrix}
0&-i\sigma_y\\
i\sigma_y&0
\end{pmatrix}
K,
\end{align}
which satisfies $P^2=1$.
We use italic (calligraphic) symbols to indicate the symmetry operator of the BdG Hamiltonian (normal-state Hamiltonian).

Let us assume that the normal state has $C_{n}$ symmetry about the $z$-axis, so that ${\cal C}_{n}h({\bf k}){\cal C}_{n}^{-1}=h(R_{n}{\bf k})$, where $R_{n}{\bf k}$ indicates the momentum after ${\cal C}_{n}$ rotation of ${\bf k}$.
When $\Delta({\bf k})$ is an eigenfunction of $C_{n}$, i.e.,
${\cal C}_{n}\Delta({\bf k}){\cal C}_{n}^{-1}=\lambda\Delta(R_{n}{\bf k})$,
$H_{\rm BdG}$ is symmetric under
$C_{n}\equiv{\rm diag}[{\cal C}_{n}, \lambda {\cal C}_{n}]$ which satisfies
$C_{n}P=\lambda PC_{n}$.
Namely, $C_{n}H_{\rm BdG}({\bf k})C_{n}^{-1}=H_{\rm BdG}(R_{n}{\bf k})$.

Now we suppose that the normal state has time reversal symmetry ${\cal T}h(-{\bf k}){\cal T}^{-1}=h({\bf k})$, under ${\cal T}=i\sigma_yK$.
When the pairing function is also time-reversal-symmetric, i.e.,
${\cal T}\Delta({\bf k}){\cal T}^{-1}
=\Delta(-{\bf k})$, the BdG Hamiltonian is symmetric under $T={\rm diag}({\cal T},{\cal T})$.
Consistency with ${\cal C}_{n}$ invariance requires
$\lambda=\lambda^*$ for $\cal{T}$-preserving pairing,
because ${\cal T}{\cal C}_n={\cal C}_n{\cal T}$.
Accordingly, $C_{n}P=\pm PC_{n}$ in $T$-symmetric superconductors, such that the pairing is either even-$C_{n}$ ($\lambda=1$) or odd-$C_n$ ($\lambda=-1$).
For our analysis below, it is convenient to define the chiral symmetry operator $S=iTP$ satisfying $C_{n}S=\pm SC_{n}$ and $S^2=1$.
The commutation relations shown here are generally valid, independent of basis choice.

{\it Higher-spin Majorana fermions (HSMFs).---}
Let us first consider the surface states of TSCs with even-$C_{n}$ pairing characterized by the relation $C_{n}P=PC_{n}$, focusing on the $n=2$ case.
On a $C_{2}$-invariant surface, the simplest form of the surface states is the twofold degenerate MF with a linear dispersion relation, protected by chiral symmetry.
More explicitly, when we take the representation $S=\sigma_z$ and $T=i\sigma_yK$ with Pauli matrices $\sigma_{x,y,z}$, a Majorana surface state can be described by the Hamiltonian $H_s(k_x,k_y)=v_xk_x\sigma_x+v_yk_y\sigma_y$, which carries a winding number $w={\rm sgn}(v_xv_y)=\pm 1$, where $w=(i/4\pi)\oint_\ell d{\bf k}\cdot{\rm Tr}\left[SH_s^{-1}\nabla_{\bf k}H_s\right]$ is defined along a loop $\ell$ surrounding the node at ${\bf k}=0$.
The total winding number of surface MFs is protected by chiral symmetry, so it is robust independent of $C_{2}$ symmetry.

\begin{figure}[t]
\includegraphics[width=8.5cm]{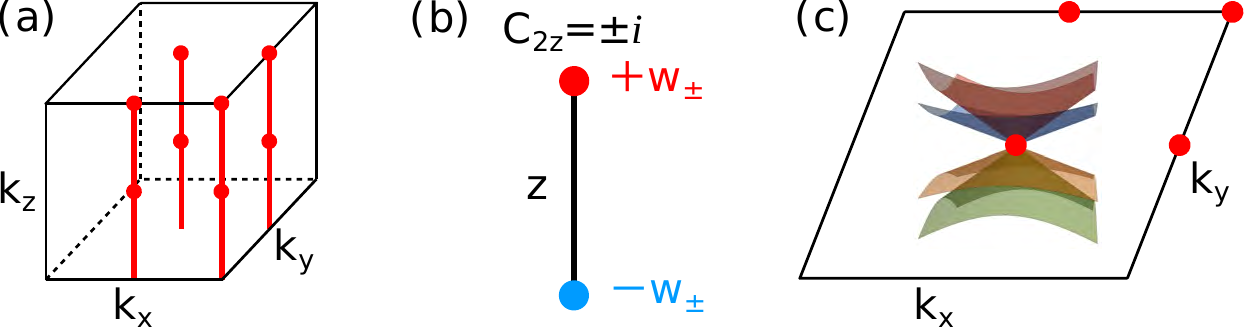}
\caption{
Bulk topology and surface higher-spin Majorana fermion.
(a) $C_{2z}$-invariant lines in the 3D Brillouin zone, which are located at $(k_x,k_y)=(0,0)$, $(\pi,0)$, $(0,\pi)$, and $(\pi,\pi)$, respectively.
(b) The 1D winding number and chirality of edge states in a $C_{2}$-invariant line.
When the winding numbers $w_{\pm}$ in the $C_2$ eigensector with eigenvalues $\pm i$ are  nonzero, $|w_{\pm}|$ Majorana zero modes appear at both edges.
The sum of the chirality (the eigenvalue of the chiral operator $S$) of the zero modes is $+w_{\pm}$ on one edge and $-w_{\pm}$ on the other edge.
(c). Spin-3/2 fermion appearing at $(k_x,k_y)=(0,0)$ of a $C_{2}$-symmetric surface Brillouin zone.
Its fourfold degeneracy originates from the nontrivial winding numbers $w_{+}=-w_{-}=\pm 2$ of the line at $(k_x,k_y)=(0,0)$.
}
\label{fig:HSF}
\end{figure}

To obtain surface states that require $C_{2}$ symmetry for their protection, let us consider a four-band surface Hamiltonian describing two overlapping MFs with opposite winding numbers: $H_s(k_x,k_y)=k_x\sigma_x+k_y\rho_z\sigma_y$, which is invariant under $S=\sigma_z$ and $T=i\sigma_yK$.
Then, possible $C_{2}$ representations commuting with $S$ and $T$, and satisfying $(C_{2})^2=-1$, are $C_{2}=-i\rho_z\sigma_z$ and $C_{2}=-i\sigma_z=-iS$.
In the former case, a mass term $m\rho_y\sigma_y$ opens the gap on the surface. 
On the other hand, in the latter case, no mass term is allowed, so the gapless spectrum is protected.
In fact, the fourfold degeneracy at ${\bf k}=0$ is {\it enforced} by the representation $C_{2}=\pm iS$ because
\begin{align}
H_s({\bf k})=-SH_s({\bf k})S^{-1}=-C_{2}H_s({\bf k})C_{2}^{-1}=-H_s(-{\bf k}),
\end{align}
so that $H_s({\bf k}={\bf 0})=0$. 
This type of symmetry-enforced degeneracy is possible only on the surface of a TSC because $C_{2}=\pm iS$ is an anomalous representation that mixes the particle-hole indices, which is impossible in an ordinary $C_2$ representation of the bulk states.

The fourfold degenerate point disperses like spin-3/2 fermions~\cite{kennett2011birefringent,bradlyn2016beyond,liang2016semimetal,tang2017multiple} because the degeneracy is lifted away from the $C_{2}$-invariant momentum ${\bf k}=0$.
In fact, the representation $C_{2}=\pm iS$ can generally protect $2n$-fold degenerate points with an arbitrary natural number $n$,
which we call spin-$(2n-1)/2$ MFs (or more generally, HSMFs).

We can understand the corresponding 3D bulk topology of $H_{\rm BdG}$ using the 1D topology on $C_{2}$-invariant lines [Fig.~\ref{fig:HSF}(a)], as shown in Ref.~\cite{fang2017topological}. 
From this, the origin of the anomalous representation $C_{2}=\pm iS$ on the surface can be found.
Let us recall that, in 1D systems with winding number $w$, $w$ zero modes with positive (negative) chirality appear on one (the other) edge~\cite{su1979solitons,chiu2016classification} [see Fig.~\ref{fig:HSF}(b)].
On $C_{2}$-invariant lines, the winding numbers $w_{\pm}$ can be defined in two distinct sectors with $C_{2}$ eigenvalues $\pm i$, respectively.
As time reversal symmetry imposes that $w_++w_-=0$~\cite{supp}, $w_+=-w_-\in {\mathbb Z}$ is the remaining invariant on a $C_2$-invariant line, which naturally leads to the anomalous representation $C_{2}=\pm iS$ at its edge.
This guarantees the protection of degeneracies at the $C_{2}$-invariant momentum on the top and bottom surfaces, as shown in Fig.~\ref{fig:HSF}(c).
As the total winding number is zero, the degeneracy at zero energy is lifted away from the $C_{2}$-invariant momentum.
Similarly, HSMFs in TSCs with even-$C_{n=3,4,6}$ pairing can be protected by the 1D winding number defined in each $C_{n}$ eigensector~\cite{supp}.

{\it Doubly charged Majorana Fermi lines (DCMFLs).---}
Next, we consider odd-$C_2$ pairing characterized by $C_{2}S=-SC_{2}$.
Odd-$C_6$ pairing also falls into this category.
In these cases, no HSMF is allowed~\cite{supp}.
Instead, surface states appear at generic momenta.
Since ${\bf k}$-local symmetries $C_{2}T$ and $C_{2}P$ satisfy $(C_{2}T)^2=(C_{2}P)^2=1$, gap nodes appear as lines, i.e., Majorana Fermi lines (MFLs), at generic momenta~\cite{bzdusek2017robust}.

\begin{figure}[t]
\includegraphics[width=8.5cm]{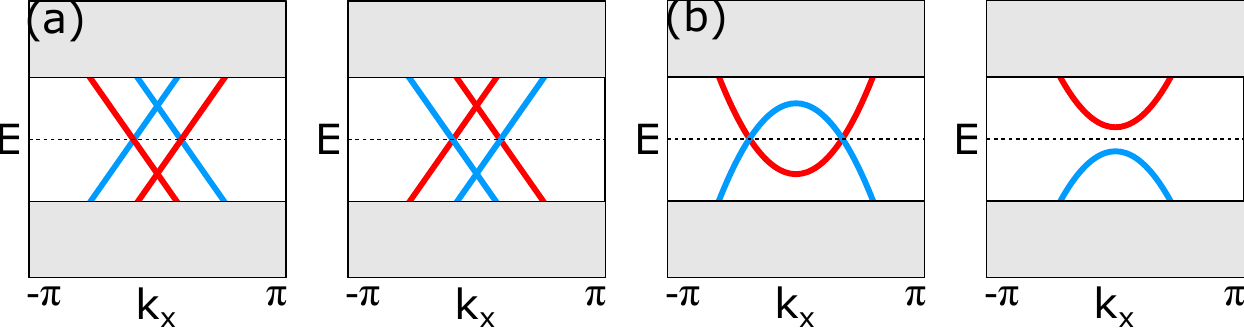}
\caption{
Single Majorana Fermi line on a $C_{2}$-invariant surface of superconductors with odd-$C_2$ pairing.
Shaded regions indicate the bulk energy spectrum of the BdG Hamiltonian.
Red and blue lines originate from the electron and hole surface bands, respectively.
The spectrum is shown along the $k_x$ direction, but the spectrum looks similar along the $k_y$ direction.
(a) Surface band structure near a DCMFL. 
(b) Surface band structure a MFL with trivial 1D charge.
Different stabilities of MFLs in (a) and (b) can be understood from the stability of the parent normal-state Fermi surfaces corresponding to the red line.
}
\label{fig:stability}
\end{figure}

To understand the anomalous MFL, let us consider a Dirac fermion on a $C_{2}$-invariant surface of a TI described by the Hamiltonian $h_{D}=-\mu+k_x\sigma_x+k_y\sigma_y$ invariant under ${\cal C}_{2}=-i\sigma_z$ and ${\cal T}=i\sigma_yK$.
Then, there is a unique odd-$C_{2}$ pairing function $\Delta(k_x,k_y)=[{\bm \Delta}\cdot {\bf k}+O(k^3)]\sigma_z$ that gives the following surface BdG Hamiltonian
\begin{align}
\label{eq:dMFL}
H_{\rm BdG}(k_x,k_y)=k_x\tau_z\sigma_x+k_y\tau_z\sigma_y
-\mu\tau_z+{\bm \Delta\cdot {\bf k}}\tau_x\sigma_z,
\end{align}
which is symmetric under $C_{2}=-i\tau_z\sigma_z$, $T=i\sigma_yK$, and $P=\tau_y\sigma_yK$ where $\tau_{x,y,z}$ are the Pauli matrices for particle-hole indices.
Gap does not open at zero energy, and an MFL appears along $|{\bf k}|=\sqrt{\mu^2+({\bm \Delta}_0\cdot {\bf k})^2}$ [Fig.~\ref{fig:stability}(a)].
The MFL does not disappear by tuning $\mu$ and $\Delta_0$, and, in fact, by any continuous deformations preserving the bulk gap.
Therefore, a single MFL of this type can appear as the characteristic surface state of odd-$C_{2}$ TSCs.

The stability of the above MFL is due to its two ${\bb Z}_2$ charges.
If we choose a basis in which $S={\rm diag}[1_{N\times N},-1_{N\times N}]$ and $C_{2}T=K$,
\begin{align}
\label{eq:BDI-Ham}
H_{\rm BdG}({\bf k})
=
\begin{pmatrix}
0& O({\bf k})\\
O^{\dagger}({\bf k}) & 0
\end{pmatrix},
\end{align}
where $O({\bf k})$ is real-valued.
The 0D topological charge is defined by the sign change of $\det O({\bf k})$ across the MFL, while the 1D topological charge is defined by the winding number of the matrix $O({\bf k})$ around a loop surrounding the MFL~\cite{bzdusek2017robust,kawakami2019topological}
Since $O({\bf k})\xrightarrow{\Delta\rightarrow0} h({\bf k})$, the topological stability of an MFL is inherited from the topological property of the normal-state Fermi surface.
The nontrivial 0D charge guarantees that a small perturbation does not gap the Fermi surface and is common to all Fermi surfaces (thus to all MFLs).
On the other hand, only the MFL arising from the Fermi surface of a single Dirac fermion carries a nontrivial 1D topological charge (inherited from the $\pi$ Berry phase of a Dirac fermion) and is robust against any continuous deformations [ Fig~\ref{fig:stability}].

Since a DCMFL is realized by odd-$C_{n=2,6}$ pairing, it accompanies gapless hinge states between side surfaces, as shown in the middle panel of Fig.~\ref{fig:scheme}(b,d).
These hinge states can be understood in terms of the $p$-wavelike (for odd-$C_2$ pairing) or $f$-wavelike (for odd-$C_6$ pairing) symmetry of the pairing function in real space: when the pairing function changes sign on the side surfaces, gapless hinge states appear as domain wall states~\cite{supp}.

{\it Quartet of Majorana fermions (QMF).---}
We again consider the surface Dirac fermions of TIs, but with an odd-$C_{4}$ pairing function.
As $(C_{2}T)^2=1$ and $(C_{2}P)^2=-1$, nodes now appear as points at generic momenta~\cite{bzdusek2017robust}.
There are two possible $C_{4}$ representations for odd-$C_4$ pairing: $C_{4z}=i\tau_x\sigma_ze^{-i\frac{\pi}{4}\sigma_z}$ and $C_{4z}=\tau_ze^{-i\frac{\pi}{4}\sigma_z}$.
In both cases, the pairing term has the form
\begin{align}
\delta H_{\Delta}
=\Delta_1({\bf k})\tau_x+\Delta_2({\bf k})\tau_x\sigma_x+\Delta_3({\bf k})\tau_x\sigma_y,
\end{align}
where the $\Delta_1\tau_x$ term is a potential mass term that anticommutes with the Dirac Hamiltonian $\tau_z\otimes h_{D}$.
The representation $C_{4z}=i\tau_x\sigma_ze^{-i\frac{\pi}{4}\sigma_z}$ allows a mass term $\Delta_1=m$, thus giving trivial surface states.
On the other hand, $C_{4z}=\tau_ze^{-i\frac{\pi}{4}\sigma_z}$ forbids such a constant mass term.
In this case, pairing terms split the fourfold degeneracy at $\bf{k}=0$ into four MFs with twofold degeneracy, as shown in Fig.~\ref{fig:scheme}(c).

In contrast to HSMFs and DCMFLs, $P$ and $S$ symmetries do not play a critical role in the protection of the QMF, so the surface structure is similar to that in TIs with $C_n$ symmetry~\cite{fang2019new}.
While the presence of $S$ symmetry promotes the ${\mathbb Z}_2$-valued Berry phase of each twofold degenerate MF to the integer-valued winding number, 
the stability of the MFs as a whole still has a ${\mathbb Z}_2$ character~\cite{supp}.
In the case of odd-$C_{4}$ pairing, as the winding numbers of MFs related by $C_{4}$ symmetry have opposite signs~\cite{supp}, the total winding number of all MFs is zero.
However, MFs carry another ${\mathbb Z}_2$ topological charge instead, indicating their stability when they merge at a $C_{4}$-invariant momentum~\cite{supp,fang2019new}.
A QMF is robust because this ${\mathbb Z}_2$ charge is nontrivial. 

Similar to the case of DCMFL, the QMF accompanies gapless hinge states between gapped side surfaces, as shown in Fig.~\ref{fig:scheme}(c).
The appearance of hinge states can be attributed to the $d$-wavelike symmetry of the pairing function in real space~\cite{supp}.

{\it Lattice model.---}
To demonstrate our theory, we consider the following model Hamiltonian describing
a doped ${\mathbb Z}_2$ TI or Dirac semimetal,
\begin{align}
\label{eq:modelH1}
h_1
&=-\mu+(4-2\cos k_x-2\cos k_y-\cos k_z)\rho_z+\sin k_x\rho_x\sigma_z\notag\\
&-\sin k_y\rho_y+(3\sin k_z(\cos k_y-\cos k_x)+m_0 \sin k_z)\rho_x\sigma_x\notag\\
&+(-\sin k_z\sin k_x\sin k_y+m_1 \sin k_z)\rho_x\sigma_y,
\end{align}
where $\rho_{i=x,y,z}$ and $\sigma_{i=x,y,z}$ are Pauli matrices for orbital and spin degrees of freedom, respectively.
This is symmetric under time reversal ${\cal T}=i\sigma_y K$ and 
mirror operations ${\cal M}_x=i\sigma_x$, ${\cal M}_y=i\rho_z\sigma_y$, and ${\cal M}_z=i\sigma_z$.
This model describes a $C_{4z}$ symmetric Dirac semimetal when $m_0=m_1=0$. Nonzero $m_0$ and $m_1$, which breaks $C_{4z}$, $M_x$ and $M_y$ symmetries, opens a gap at bulk Dirac points leading to a ${\mathbb Z}_2$ TI~\cite{kobayashi2015topological,hashimoto2016superconductivity}.

We first consider even-$C_{2z}$ pairing.
When $\mu$ is small, as the ${\bf k}=(0,0,k_z)$ line is the only $C_2$-invariant line crossing the Fermi surface,
we need a 1D TSC with $w_{+}=-w_-=2$ along the ${\bf k}=(0,0,k_z)$ line to observe a HSMF on the boundary.
If we choose a pairing function $\Delta({\bf k})=\Delta_e\sin k_z\sigma_z$, the resulting BdG Hamiltonian along the ${\bf k}=(0,0,k_z)$ line is 
$H_{\rm BdG}=-\mu\tau_z-\cos k_z\rho_{z}\tau_z+\Delta_e \sin k_z \sigma_{z}\tau_x$, where $\tau_{i=x,y,z}$ are Pauli matrices for particle-hole indices, and $m_0=m_1=0$ is assumed for simplicity.
From this 1D Hamiltonian, we obtain $w_{+}=-w_{-}=2$ for $C_{2z}=-i\rho_z\sigma_z$~\cite{supp}.
The associated surface spectrum with a spin-3/2 fermion at $(k_x,k_y)=(0,0)$ on the $C_{2z}$-invariant surface is shown in Fig.~\ref{fig:numerical}(a).
When other even-$C_{2}$ pairing terms dominate, nodal or topologically trivial superconductors can also be obtained~\cite{supp}.

On the other hand, in the case of odd-$C_{2}$ pairing, as the presence of a single Dirac fermion on the surface is key for observing a DCMFL, any odd-$C_{2}$ pairing can induce a DCMFL, thus realizing a $C_{2}$-protected TSC as long as the bulk gap fully opens.
The surface spectrum for $\Delta({\bf k})=\Delta_o \rho_x$ is shown in Fig.~\ref{fig:numerical}(b).
Here, we need both $m_0$ and $m_1$ to be nonzero to obtain a fully gapped TSC; otherwise, bulk Dirac points protected by either $C_{4}$ symmetry~\cite{kobayashi2015topological,hashimoto2016superconductivity} or mirror symmetry~\cite{yang2014dirac} appear.
In addition to DCMFLs, we obtain helical Majorana hinge states on side surfaces [Fig.~\ref{fig:numerical}(c)].

To describe an odd-$C_4$ TSC with QMF, 
we need a model whose Fermi surface does not cross $C_{4}$-invariant lines; otherwise, the bulk gap does not fully open for odd-$C_{4}$ pairing.
Hence, instead of Eq.~(\ref{eq:modelH1}), we consider the following model Hamiltonian for a doped ${\mathbb Z}_2$ TI, 
\begin{align}
\label{eq:modelH2}
h_{2}=&-\mu +\sin k_z\rho_y + (M-\sum_{i=x,y,z}\cos k_i)\rho_z
\nonumber\\
&+\lambda_{{\rm SO}}(\sin k_x\rho_x\sigma_y-\sin k_y\rho_x\sigma_x),
\end{align}
where
$\lambda_{{\rm SO}}$ indicates spin-orbit coupling.
$h_2$ is symmetric under ${\cal T}=i\sigma_yK$, ${\cal M}_x=i\sigma_x$, ${\cal M}_y=i\sigma_y$, ${\cal M}_z=i\rho_z\sigma_z$, and ${\cal C}_{4z}=e^{-i\frac{\pi}{4}\sigma_z}$.
If we take $|\mu|$ larger than the gap induced by $\lambda_{{\rm SO}}$, the system has a torus-shaped Fermi surface, which is a characteristic of nodal line semimetals.
As this Fermi surface does not cross a $C_{4z}$-invariant line, a fully gapped TSC can be obtained by introducing an odd-$C_{4z}$ pairing.
If we consider $\Delta_d({\bf k})=\Delta_{x^2-y^2}(\cos k_y-\cos k_x)\rho_y\sigma_z+\Delta_{xy}\sin k_x\sin k_y\rho_x+\Delta_{\delta}(\sin k_x\rho_x\sigma_y+\sin k_y\rho_x\sigma_x)$, the bulk and side-surface gap fully opens and QMF (hinge states) appears on the top surface (side hinges) [Fig.~\ref{fig:numerical}(d,e)].

\begin{figure}[t!]
\includegraphics[width=8.5cm]{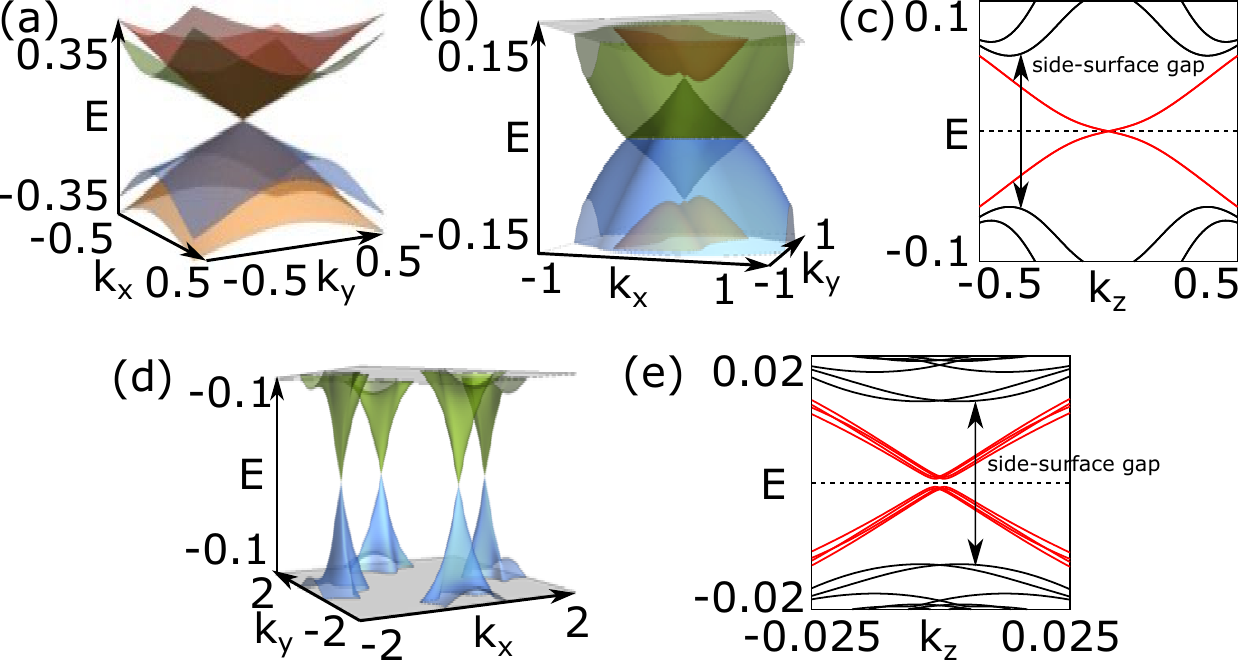}
\caption{
Boundary Majorana surface states in lattice models.
(a) Spin-3/2 MF obtained from Eq.~\eqref{eq:modelH1} with $\mu=0.5$, $m_0=m_1=0$ and even-$C_2$ pairing $\Delta_e=0.5$.
(b) A DCMFL on a $C_2$ invariant surface obtained from Eq.~\eqref{eq:modelH1} with $\mu=0.5$, $m_0=m_1=0.3$ and odd-$C_2$ pairing $\Delta_o=0.3$.
(c) The helical hinge state (red color) for odd-$C_2$ pairing obtained from the same Hamiltonian used in (b). 
(d) A quartet of spin-1/2 MFs obtained from Eq.~\eqref{eq:modelH2} with $\mu=0.2$, $M=1.5$, $\lambda_{\rm SO}=0.1$, $\Delta_{x^2-y^2}=\Delta_{xy}=0.5$, and $\Delta_{\delta}=0.1$.
Associated hinge states (red color) are shown in (e).
The splitting of the four $C_{4z}$-related hinge spectra and small gap at zero energy are finite-size effects, which decrease exponentially as the size of the system grows.
The energy spectra in (a), (b), and (d) are calculated with $40$ unit cells along the $z$ direction and periodic boundary conditions along the $x$, $y$ directions.
(c) and (e) are calculated with $20\times 20$ unit cells along the $x$ and $y$ directions, and periodic boundary conditions along the $z$ direction.
}
\label{fig:numerical}
\end{figure}

\section{Discussion}

Our model study shows that doped $Z_2$ TIs having a band structure of massive Dirac semimetals are promising candidates for rotation-protected TSCs.
Au$_{2}$Pb is such a material~\cite{schoop2015dirac,xing2016superconductivity,wu2018electronic}.
It has an orthorhombic symmetry and shows a fully gapped superconductivity below 1.2 K~\cite{schoop2015dirac}.
While this system has been proposed as a TSC, this cannot be a Fu-Kane $Z_2$ TSC because it does not have Fermi surfaces enclosing a time-reversal-invariant momentum~\cite{fu2010odd,sato2009topological,sato2010topological,xing2016superconductivity}.
On the other hand, it is more likely that Au$_2$Pd is a rotation-protected TSC hosting either HSMF or DCMFL.
Detailed experimental studies on pairing symmetry and superconducting surface spectrum are desired to test the scenario we propose.

Optical responses may be able to distinguish HSMF and DCMFL because DCMFL can show subgap optical responses down to zero photon energy while other MFs show zero optical response~\cite{ahn2021theory}.
Finding other characteristic physical responses of unconventional MFs will be a promising future direction.

Interaction and disorder effects on new MFs can be an interesting subject.
In the case of HSMFs, the ${\bb Z}$ classification will reduce to ${\bb Z}_8$, because the winding number in each eigenspace will take a ${\bb Z}_8$ value in interacting systems~\cite{fidkowski2010effects,fidkowski2011topological,turner2011topological,you2014topological}.
It is an open question whether further modification of the classification will occur.
Also, while crystalline-symmetry-protected states are stable against averaged disorder~\cite{fu2012topology,hsieh2012topological,fang2015new}, the fate of them under strong disorder needs to be studied further.

\begin{acknowledgments}
J.A. thanks SangEun Han and James Jun He for helpful discussions.
J.A. was supported by IBS-R009-D1.
B.J.Y. was supported by the Institute for Basic Science in Korea (Grant No. IBS-R009-D1),
Samsung Science and Technology Foundation under Project Number SSTF-BA2002-06,
Basic Science Research Program through the National Research Foundation of Korea (NRF) (Grant No. 0426-20210011),
and the U.S. Army Research Office and and Asian Office of Aerospace Research \& Development (AOARD) under Grant Number W911NF-18-1-0137.
\end{acknowledgments}


%

\clearpage
\newpage

\onecolumngrid

\begin{center}
\textbf{\large Supplemental Material for ``Unconventional Majorana fermions on the surface of topological superconductors protected by rotational symmetry"}
\end{center}

\begin{center}
Junyeong Ahn$^{1,2,3,4,5,*}$ and Bohm-Jung Yang$^{1,2,3,\dagger}$\\
\vspace{0.1in}
{\small \it
$^1$Center for Correlated Electron Systems, Institute for Basic Science (IBS), Seoul 08826, Korea\\
$^2$Department of Physics and Astronomy, Seoul National University, Seoul 08826, Korea\\
$^3$Center for Theoretical Physics (CTP), Seoul National University, Seoul 08826, Korea\\
$^4$RIKEN Center for Emergent Matter Science (CEMS), Wako, Saitama 351-0198, Japan\\
$^5$Department of Applied Physics, The University of Tokyo, Bunkyo, Tokyo 113-8656, Japan}
\end{center}

\setcounter{section}{0}
\setcounter{equation}{0}
\setcounter{figure}{0}
\setcounter{table}{0}
\setcounter{page}{1}

\tableofcontents

\newpage

\twocolumngrid

\section{Classification of $C_{n}$-protected topological superconductors}
\label{sec:classification}
Let us classify topological superconductors protected by $C_{n}$ symmetries.
Before we move on to the topological classification, we note that the pairing function should be an eigenfunction of $C_{n}$ in order to preserve the $C_{n}$ symmetry in the superconducting phase (more precisely, to preserve the symmetry under the combination of $C_{n}$ and a global phase rotation).
Since $C_{n}$ and $C$ symmetries do not protect a nodal point or a nodal loop on the surface without time reversal symmetry, our discussion below is concentrated on the time-reversal-symmetric superconductors.
To keep time reversal symmetry as well as $C_{n}$ symmetry in the superconducting phase, the pairing function has a real eigenvalue under the $G$ operation.
Otherwise, the pairing function is not invariant under time reversal symmetry because time reversal changes the eigenvalue of the pairing function.
Thus, it is enough to consider only even- or odd-$C_{n}$ pairing ${\cal C}_{n}\Delta({\bf k}){\cal C}_{n}^{-1}=\pm \Delta(R_{n}{\bf k})$ for time-reversal-symmetric superconductors.

Throughout the classification, we assume that $C_{n}$ commutes with $T$, i.e., $C_{n}T=TC_{n}$.
It is satisfied in physically relevant systems such as nonmagnetic systems with finite/negligible spin-orbit coupling (class DIII with $C_{n}^n=-1$ and class CI with $C_{n}^n=1$) and spin-polarized systems without spin-orbit coupling (class BDI with $C_{n}^n=1$).

\begin{table}[t!]
\begin{tabular}{ccc|cc}
$(C_{2}T)^2$ 	& $(C_{2}P)^2$	& $S^2$	&0D	&1D\\
\hhline{===|==}
$0$		&$0$		&$0$
&$0$		&$0$\\
$0$		&$0$		&$1$
&$0$		&${\bb Z}_{\rm U(1)}$\\
\hline
$1$		&$1$		&$1$
&${\bb Z}_2$		&${\bb Z}_2$\\
$1$		&$-1$		&$1$
&$0$		&${\bb Z}_{\rm U(1)}$\\
$-1$		&$1$		&$1$
&$0$		&$2{\bb Z}_{\rm U(1)}$\\
$-1$		&$-1$		&$1$
&$0$		&$0$\\
\end{tabular}
\caption{
Topological charges of gapless nodes at a generic momentum in the surface Brillouin zone.
$C_{2}$ is a twofold rotation around the surface normal axis.
$C_{2z}T$, $C_{2}P$, and $S$ do not change crystal momentum, so they constrain the band topology and gap closing condition at generic momenta.
${\bb Z}_{\rm U(1)}$ is the U(1) winding number carried by point nodes, 0D ${\bb Z}_{2}$ is the topological charge protecting line nodes, and 1D ${\bb Z}_{2}$ is the O(N) winding number that carried by a line node as a secondary topological charge~\cite{bzdusek2017robust}.
}
\label{tab:charge}
\end{table}

\begin{table*}[t!]
\begin{tabular}{c|cccc|cccc|cc|c}
AZ class		& $(C_{n})^n$		&$T^2$ 	& $P^2$	& $S^2$	&even-$C_{2}$	&even-$C_{3}$	&even-$C_{4}$	& even-$C_{6}$ & odd-$C_{n=2,6}$	& odd-$C_{4}$	& complex-$C_{n}$\\
\hhline{=|====|====|===}
DIII			 	
&$-1$		& $-1$		&$1$		&$1$
&${\bb Z}_{\rm U(1)}\times {\bb Z}_{\rm HS}$	
&${\bb Z}_{\rm U(1)}\times {\bb Z}_{\rm HS}$
&${\bb Z}_{\rm U(1)}\times {\bb Z}_{\rm HS}^2$ 
& ${\bb Z}_{\rm U(1)}\times {\bb Z}_{\rm HS}^3$
&$({\bb Z}_2)_{\rm DC}$	&$({\bb Z}_2)_{\rm M}$
&$0$\\
CI						 
&$-1$		& $1$		&$-1$			&$1$
&$2{\bb Z}_{\rm U(1)}\times {\bb Z}_{\rm HS}$	
&$2{\bb Z}_{\rm U(1)}\times {\bb Z}_{\rm HS}$
&$2{\bb Z}_{\rm U(1)}\times {\bb Z}_{\rm HS}^2$ 
&$2{\bb Z}_{\rm U(1)}\times {\bb Z}_{\rm HS}^3$
&$0$	&$0$
&$0$\\
CII	
&$-1$		&$-1$		&$-1$		&$1$
&$({\bb Z}_2)_{\rm DC}$
&${\bb Z}_{\rm HS}$
&$({\bb Z}_2)_{\rm DC}\times{\bb Z}_{\rm HS}$
&$({\bb Z}_2)_{\rm DC}\times{\bb Z}_{\rm HS}^2$	
&$({\bb Z}_2)_{\rm M}$	&$({\bb Z}_2)_{\rm DC}$
&$0$\\
BDI	
&$-1$		&$1$		&$1$		&$1$
&$0$
&${\bb Z}_{\rm HS}$
&${\bb Z}_{\rm HS}$
&${\bb Z}_{\rm HS}^2$		
&$0$	&$0$
&$0$\\
\hline
DIII			
&$1$		& $-1$		&$1$			&$1$
&${\bb Z}_{\rm U(1)}$
&${\bb Z}_{\rm U(1)}\times {\bb Z}_{\rm HS}$
&${\bb Z}_{\rm U(1)}\times {\bb Z}_{\rm HS}$
&${\bb Z}_{\rm U(1)}\times {\bb Z}_{\rm HS}^2$
&$0$		&$0$
&$0$\\
CI			
&$1$		& $1$		&$-1$			&$1$
&$2{\bb Z}_{\rm U(1)}$
&$2{\bb Z}_{\rm U(1)}\times {\bb Z}_{\rm HS}$
&$2{\bb Z}_{\rm U(1)}\times {\bb Z}_{\rm HS}$
&$2{\bb Z}_{\rm U(1)}\times {\bb Z}_{\rm HS}^2$
&$0$		&$0$
&$0$\\
CII	 	
&$1$		&$-1$		&$-1$		&$1$
&${\bb Z}_{\rm HS}$
&${\bb Z}_{\rm HS}$
&${\bb Z}_{\rm HS}^2$ 
&${\bb Z}_{\rm HS}^3$	
&$0$	&$0$
&$0$\\
BDI	 	
&$1$		&$1$		&$1$		&$1$
&${\bb Z}_{\rm HS}$
&${\bb Z}_{\rm HS}$
&${\bb Z}_{\rm HS}^2$ 
&${\bb Z}_{\rm HS}^3$	
&$0$	&$0$
&$0$\\
\hline
C or D				 		
&$\pm 1$		&$0$		&$\pm 1$		&$0$
&$0$			&$0$			&$0$			&$0$
&$0$		&$0$
&$0$\\
\hline
AIII						 		
&$\pm 1$		&$0$		&$0$		&$1$
&${\bb Z}_{\rm U(1)}\times {\bb Z}_{\rm HS}$
&${\bb Z}_{\rm U(1)}\times {\bb Z}_{\rm HS}^2$
&${\bb Z}_{\rm U(1)}\times {\bb Z}_{\rm HS}^3$ 
&${\bb Z}_{\rm U(1)}\times {\bb Z}_{\rm HS}^5$
&$0$		&$0$
&$0$
\end{tabular}
\caption{
Topological classification of $C_{n}$-symmetric 3D superconductors.
$T$, $P$, and $S$ represent time reversal, particle-hole, and chiral symmetries used for Altland-Zirnbaur (AZ) symmetry classification.
We assume $C_{n}T=TC_{n}$ and define $T$ and $P$ such that $TP=PT$.
$C_{n}$ and $P$ satisfies the commutation relation $C_{n}P=\lambda PC_{n}$.
Even- and odd-$C_{n}$ correspond to $\lambda=+1$ and $\lambda=-1$, respectively, and complex-$C_{n}$ indicate that $\lambda$ is complex-valued.
The subscript $\rm U(1)$ indicates that it is the 3D winding number of the Hamiltonian protected by chiral $S$ symmetry.
The subscripts $\rm HS$, $\rm DC$, and $\rm M$ indicate that they are responsible for the protection of higher-spin Majorana fermions (HSMFs), doubly charged Majorana Fermi lines (DCMFLs), and a $C_{n}$-multiplet of Majorana fermions (MMF), respectively, on the $C_{n}$-preserving surfaces.
When time reversal symmetry is broken, there is no 3D topological superconductor protected by $C_{n}$ symmetry.
In class AIII, even-, odd-, and complex-$C_{n}$ indicate the commutation relation $C_{n}S=\lambda SC_{n}$ instead of $C_{n}P=\lambda PC_{n}$.
}
\label{tab:symmetry}
\end{table*}

\subsection{Even-$C_{n}$ ($C_{n}$-invariant) superconducting pairing}

To study the protection of gapless states on the surface, we need to consider both $C_{n}$-invariant momenta and generic momenta.
At generic momenta, there is no $C_{n}$-protected gapless point nodes for even-$C_{n}$ pairing because $C_{n}$-related gapless states have the same U(1) winding number, so they are robust against breaking $C_{n}$ symmetry as long as chiral symmetry is preserved and are thus not $C_{n}$-protected states.
Therefore, only line nodes can appear as anomalous surface states at generic momenta.
Since the 1D topological charge of line nodes is protected by rotational symmetries and is not the U(1) winding number protected by chiral symmetry [Table.\ref{tab:charge}], line nodes can appear as $C_{n}$-protected anomalous surface states.

As for the high-symmetry momenta, it is enough to analyze $C_{n}$-invariant momenta.
For other high-symmetry momenta that are invariant under a subgroup of the $C_{n}$ group, the representation of the subgroup at $C_{n}$-invariant momenta already captures the related topology.
For example, let us consider the $C_{2}$-invariant surface momenta $\bar{X}=(\pi,0)$ and $\bar{Y}=(0,\pi)$ of a $C_{4}$-symmetric system.
The $C_{2}$-protected higher-spin spectrum occuring at $\bar{X}$ and $\bar{Y}$ also occur at $\bar{\Gamma}=(0,0)$ and $\bar{M}=(\pi,\pi)$.
Since we consider strong topological phases that are robust against translation symmetry breaking that preserves $C_{n}$ symmetry, we can consider a unit cell doubling along both $x$ and $y$ directions (such that four unit cells merge into a large unit cell).
By this process, the higher-spin spectrum at $\bar{X}$ and $\bar{Y}$ are folded into the $\bar{\Gamma}$ point, but the strong topological phase does not change by definition.
Therefore, the analysis of at the surface $\bar{\Gamma}$ point is enough to study the 3D strong topological phase protected by $C_{4}$ symmetry.
We thus focus on the $C_{n}$-invariant line in the 3D Brillouin zone, the 1D winding number on which is responsible for the protection of the gapless states at the corresponding $C_{n}$-invariant momentum (as shown in the main text for $C_{2}$).

In the following, we first classify the higher-spin Majorana fermions at $\bar{\Gamma}$, i.e., classify 1D winding numbers along the line $(0,0,k_z)$ in the 3D bulk Brillouin zone.
The classification of the 1D winding numbers for two physically relevant classes of spin-orbit coupled systems and spin-SU(2)-symmetric systems were done in Ref.~\cite{fang2017topological}.
We extend this to other symmetry classes.
After that, we identify symmetry classes that host doubly charged Majorana Fermi lines.

\subsubsection{Chiral-symmetric systems without time reversal symmetry (class AIII)}

Let us first forget about time reversal symmetry and consider $C_{n}$ and chiral symmetries only.
In the case of $C_{2}$ symmetry, we have two 1D invariants for each eigensector
\begin{align}
\lambda
\in
\{\lambda_{\pm}\}
=
\begin{cases}
\{ -i,i\}&(C_2^2=-1)\notag\\
\{-1, 1\}&(C_2^2=1)
\end{cases}.
\end{align}
Since $w=w_++w_-$ is total the winding number, which serves as the 1D topological invariant of the system and is protected by chiral symmetry, there remains only one 1D invariant relevant for the 3D topology protected by the simultaneous presence of $C_{2}$ and chiral symmetries.
Similary, there are two, three, and five crystalline 1D invariants for $C_{3}$, $C_{4}$, and $C_{6}$ symmetries because there are three eigenvalues for
$C_{3}$
\begin{align}
\lambda
\in
\{\lambda_{1\pm},\lambda_{2}\}
=
\begin{cases}
\{ e^{\pm \pi i/3}, -1\}&(C_3^3=-1)\notag\\
\{e^{\pm 2\pi i/3}, 1\}&(C_3^3=1)
\end{cases},
\end{align}
four eigenvalues for $C_{4}$
\begin{align}
\lambda
&\in\{\lambda_{1\pm},\lambda_{2\pm}\}
=\{ e^{\pm \pi i/4}, e^{\pm 3\pi i/4}\}&(C_4^4=-1)\notag\\
\lambda
&\in\{\lambda_{1\pm},\lambda_{2},\lambda_{3}\}
=\{ e^{\pm \pi i/2}, -1,1\}&(C_4^4=1),
\end{align}
and six eigenvalues for $C_{6}$
\begin{align}
\lambda
&\in\{\lambda_{1\pm},\lambda_{2\pm},\lambda_{3\pm}\}
=\{ e^{\pm \pi i/6}, e^{\pm \pi i/2},e^{\pm 5\pi i/6}\},\notag\\
\lambda
&\in\{\lambda_{1\pm},\lambda_{2\pm},\lambda_{3},\lambda_{4}\}
=\{e^{\pm \pi i/3},e^{\pm 2\pi i/3},-1,1\}.
\end{align}
for $C_6^6=-1$ and $C_6^6=1$, respectively.
Thus, in systems with $C_{n}$ and $S$ symmetries, we have
\begin{align}
&{\bb Z}_{\rm U(1)}\times {\bb Z}_{\rm HSF}\text{ for $C_2$},\notag\\
&{\bb Z}_{\rm U(1)}\times {\bb Z}_{\rm HSF}^2\text{ for $C_3$},\notag\\
&{\bb Z}_{\rm U(1)}\times {\bb Z}_{\rm HSF}^3\text{ for $C_4$},\notag\\
&{\bb Z}_{\rm U(1)}\times {\bb Z}_{\rm HSF}^5\text{ for $C_6$},
\end{align}
independent of the presence of spin-orbit coupling.

\subsubsection{Constraints from time reversal symmetry}

Let us now consider the effect of time reversal symmetry in chiral-symmetric systems, described by
\begin{align}
\label{eq:1Dwinding-eig-T}
w^{\lambda}_{\rm 1D}
&=+(-1)^{s_T+s_C}w^{\lambda^*}_{\rm 1D},\notag\\
w_{\rm 3D}
&=-(-1)^{s_T+s_C}w_{\rm 3D}.
\end{align}
which we derive in Sec.~\ref{sec:winding}.
Here, $s_T$ and $s_C$ are defined by $T^2=(-1)^{s_T}$ and $P^2=(-1)^{s_C}$.
First, time reversal symmetry can impose constraints on ${\bb Z}_{\rm U(1)}$: it is always trivial when the AZ class is BDI or CII where $T^2=P^2$.
Also, in class CI, we have $2{\bb Z}_{\rm U(1)}$ instead of ${\bb Z}_{\rm U(1)}$ because spinless time reversal symmetry forbids an odd 3D winding number, as shown in Sec.~\ref{sec:Berry}.
Second, the $C_{n}$-protected invariants ${\bb Z}_{\rm HSF}$ depends on whether the representation satisfies $(C_{n})^n=1$ or $-1$.
We explicitly count the number of constraints on the ${\bb Z}_{\rm HSF}$ invariants from time reversal symmetry below.

\subsubsection{AZ classes DIII and CI}

In spin-orbit coupled systems and spin-SU(2)-symmetric systems, the AZ class is DIII and CI, respectively.
For spin-orbit coupled systems, where $(C_{n})^n=-1$, time reversal symmetry constraints are 
\begin{align}
w_{+}+w_{-}
&=0,
\end{align}
for $C_{2}$ eigensectors
\begin{align}
w_{+}+w_{-}
&=0,\notag\\
w_0
&=0
\end{align}
for $C_{3}$ eigensectors,
\begin{align}
w_{1+}+w_{1-}
&=0,\notag\\
w_{2+}+w_{2-}
&=0
\end{align}
for $C_{4}$ eigensectors,
and
\begin{align}
w_{1+}+w_{1-}
&=0,\notag\\
w_{2+}+w_{2-}
&=0,\notag\\
w_{3+}+w_{3-}
&=0
\end{align}
for $C_{6}$ eigensectors.
Here, we do not count the constraint on the total winding number $w=\sum_iw_i$, because it is irrelevant for the 3D topological phase.
Thus, by the zero, one, one, and two constraints on $C_{n=2,3,4,6}$ eigensectors from time reversal symmetry, the topological classification is reduced to
\begin{align}
&{\mathbb Z}_{\rm U(1)}\times {\mathbb Z}_{\rm HSF}\text{ for $C_2$},\notag\\
&{\mathbb Z}_{\rm U(1)}\times {\mathbb Z}_{\rm HSF}\text{ for $C_3$},\notag\\
&{\mathbb Z}_{\rm U(1)}\times {\mathbb Z}_{\rm HSF}^2\text{ for $C_4$},\notag\\
&{\mathbb Z}_{\rm U(1)}\times {\mathbb Z}_{\rm HSF}^3\text{ for $C_6$}.
\end{align}

For spin-SU(2)-rotation-symmetric systems, where $(C_{n})^n=1$,
\begin{align}
w_{+}
&=0,\notag\\
w_{-}
&=0,
\end{align}
for $C_{2}$ eigensectors
\begin{align}
w_{+}+w_{-}
&=0,\notag\\
w_0
&=0
\end{align}
for $C_{3}$ eigensectors,
and
\begin{align}
w_{1+}+w_{1-}
&=0,\notag\\
w_{2}
&=0,\notag\\
w_{3}
&=0
\end{align}
for $C_{4}$ eigensectors,
and
\begin{align}
w_{1+}+w_{1-}
&=0,\notag\\
w_{2+}+w_{2-}
&=0,\notag\\
w_{3}
&=0,\notag\\
w_{4}
&=0
\end{align}
for $C_{6}$ eigensectors.
Again, the constraint $w=0$ on the total 1D winding number does not affect the classification of 3D topological phases.
Thus, we have a classification
\begin{align}
2{\mathbb Z}_{\rm U(1)}&\text{ for $C_2$},\notag\\
2{\mathbb Z}_{\rm U(1)}\times {\mathbb Z}_{\rm HSF}&\text{ for $C_3$},\notag\\
2{\mathbb Z}_{\rm U(1)}\times {\mathbb Z}_{\rm HSF}&\text{ for $C_4$},\notag\\
2{\mathbb Z}_{\rm U(1)}\times {\mathbb Z}_{\rm HSF}^2&\text{ for $C_6$}.
\end{align}

We can also obtain a classification for the class CI systems with $(C_{n})^n=-1$ and the class DIII systems with $(C_{n})^n=1$ although their relation to physical systems is not clear.
The classifications are respectively
\begin{align}
&2{\bb Z}_{\rm U(1)}\times {\bb Z}_{\rm HSF}\text{ for $C_2$},\notag\\
&2{\bb Z}_{\rm U(1)}\times {\bb Z}_{\rm HSF}\text{ for $C_3$},\notag\\
&2{\bb Z}_{\rm U(1)}\times {\bb Z}_{\rm HSF}^2\text{ for $C_4$},\notag\\
&2{\bb Z}_{\rm U(1)}\times {\bb Z}_{\rm HSF}^3\text{ for $C_6$},
\end{align}
and
\begin{align}
{\bb Z}_{\rm U(1)}&\text{ for $C_2$},\notag\\
{\bb Z}_{\rm U(1)}\times {\bb Z}_{\rm HSF}&\text{ for $C_3$},\notag\\
{\bb Z}_{\rm U(1)}\times {\bb Z}_{\rm HSF}&\text{ for $C_4$},\notag\\
{\bb Z}_{\rm U(1)}\times {\bb Z}_{\rm HSF}^2&\text{ for $C_6$}.
\end{align}

\subsubsection{AZ classes BDI and CII}

In spin-polarized systems without spin-orbit coupling, although time reversal and spin-rotation symmetries are broken separately, the combination of time reversal and a $\pi$-rotation of spin (around the axis perpendicular to the magnetic order) defines an effective time reversal satisfying $T^2=1$, so the AZ class is BDI.
Also, a spinless representation of rotation operators is possible because spin and orbital degrees of freedom are independent, i.e., $(C_{n})^n=1$.
In this case, time reversal symmetry constraints on the 1D winding numbers are
none for $C_{2}$ eigensectors,
\begin{align}
w_{+}-w_{-}
&=0
\end{align}
for $C_{3}$ eigensectors,
\begin{align}
w_{1+}-w_{1-}
&=0
\end{align}
for $C_{4}$ eigensectors
and
\begin{align}
w_{1+}-w_{1-}
&=0,\notag\\
w_{2+}-w_{2-}
&=0
\end{align}
for $C_{6}$ eigensectors.
Time reversal symmetry also imposes that the 3D winding number vanishes.
Thus,
\begin{align}
{\bb Z}_{\rm HSF}&\text{ for $C_2$},\notag\\
{\bb Z}_{\rm HSF}&\text{ for $C_3$},\notag\\
{\bb Z}_{\rm HSF}^2&\text{ for $C_4$},\notag\\
{\bb Z}_{\rm HSF}^3&\text{ for $C_6$}.
\end{align}

For completeness, we also consider $(C_{n})^n=-1$ in the class BDI.
We have constraints
\begin{align}
w_{+}-w_{-}
&=0.
\end{align}
for $C_{2}$ eigensectors
\begin{align}
w_{+}-w_{-}
&=0
\end{align}
for $C_{3}$ eigensectors,
\begin{align}
w_{1+}-w_{1-}
&=0,\notag\\
w_{2+}-w_{2-}
&=0
\end{align}
for $C_{4}$ eigensectors
and
\begin{align}
w_{1+}-w_{1-}
&=0,\notag\\
w_{2+}-w_{2-}
&=0,\notag\\
w_{3+}-w_{3-}
&=0
\end{align}
for $C_{6}$ eigensectors.
Thus,
\begin{align}
0					&\text{ for $C_2$},\notag\\
{\bb Z}_{\rm HSF}	&\text{ for $C_3$},\notag\\
{\bb Z}_{\rm HSF}	&\text{ for $C_4$},\notag\\
{\bb Z}_{\rm HSF}^2&\text{ for $C_6$}.
\end{align}

Similarly, for class CII systems with $(C_{n})^n=1$ and $C_{n}=-1$, we have
\begin{align}
&{\bb Z}_{\rm HSF}\text{ for $C_2$},\notag\\
&{\bb Z}_{\rm HSF}\text{ for $C_3$},\notag\\
&{\bb Z}_{\rm HSF}^2\text{ for $C_4$},\notag\\
&{\bb Z}_{\rm HSF}^3\text{ for $C_6$}.
\end{align}
and
\begin{align}
0						&\text{ for $C_2$},\notag\\
{\bb Z}_{\rm HSF}		&\text{ for $C_3$},\notag\\
{\bb Z}_{\rm HSF}		&\text{ for $C_4$},\notag\\
{\bb Z}_{\rm HSF}^2	&\text{ for $C_6$},
\end{align}
respectively.

\subsubsection{Doubly charged Majorana Fermi lines}

As shown in the main text, a single doubly charged Majorana Fermi lines can appear when $T^2=-1$ and $(C_2T)^2=(C_2P)^2=1$ are satisfied ($T^2=1$ excludes the presence of a single doubly charged Majorana Fermi line).
In the case of even-$C_{n}$ pairing, the only possibility is to have $T^2=C_2^2=P^2=-1$ since we assume $C_{n}T=TC_{n}$.
Let us show that there exists a doubly charged Majorana Fermi line in this symmetry class.
We consider the following Majorana Hamiltonian
\begin{align}
H_0=
k_x\sigma_x+k_y\tau_z\sigma_y.
\end{align}
It has symmetries under $C_{2z}=i\tau_z\sigma_z$, $T=i\sigma_yK$, $P=i\tau_yK$, which satisfy $T^2=C_2^2=P^2=-1$.
Perturbations that respect these symmetries are
\begin{align}
\delta H=m_1\tau_y\sigma_x+m_2\tau_z,
\end{align}
where $m_{1,2}(-{\bf k})=m_{1,2}({\bf k})$.
Nonzero $m_{1,2}$ deforms the fourfold degenerate point to a line node, as one can see from the spectrum of
\begin{align}
\label{eq:DCMFL_even}
H=H_0+\delta H,
\end{align}
which is
\begin{align}
E=\pm \sqrt{k_x^2+k_y^2}\pm \sqrt{m_1^2+m_2^2}.
\end{align}
It shows that the gap closes at $E=0$ when $k_x^2+k_y^2=m_1^2+m_2^2$, which forms a loop generically.
The Hamiltonian $H$ respects $C_{n=2,4,6}$ symmetry with even-$C_{n}$ pairing where
\begin{align}
C_{n=2,4,6}
=e^{-i\frac{\pi}{n}\tau_z\sigma_z}.
\end{align}
$C_{n}$ symmetry requires that $m_{1,2}(R_{n}{\bf k})=m_{1,2}({\bf k})$.

\subsection{Odd-$C_{2,6}$ superconducting pairing}

Let us note that odd-$C_{6}$ pairing is even-$C_{3}$.
In this case, $C_{3}$ eigensectors have well-defined 1D winding numbers although we cannot assign the 1D winding number for $C_{6}$ eigensectors since $C_{6}$ and $S$ do not commute.
However, $C_{6}$ symmetry imposes that two sectors with eigenvalues $\lambda$ and $-\lambda$, which has the same $C_{3}$ eigenvalue, has zero winding number.
Moreover, since the odd-$C_{2,6}$ pairing also forbids a nontrivial 3D winding number, the only possibility is to have Majorana fermions at generic momenta with a trivial total winding number.

For odd-$C_{2,6}$ pairing, the surface can always be gapped when $C_{2}^2=1$ because $C_{2}$ operator itself can serve as a mass term.
To see this, we note that $C_{2}$ is a Hermitian matrix because it is a unitary matrix satisfying $C_{2}^2=1$, and it has the same symmetry property as the Hamiltonian, i.e., $(C_{2}T)C_{2}(C_{2}T)^{-1}=C_{2}$, $(C_{2}P)C_{2}(C_{2}P)^{-1}=-C_{2}$, and $(C_{n})C_{2}(C_{n})^{-1}=C_{2}$.
We thus consider the cases with $C_{2}^2=-1$ only.

\subsubsection{Doubly charged Majorana Fermi lines}

A single doubly charged Majorana Fermi line can ppear when $T^2=-1$ and $(C_2T)^2=(C_2P)^2=1$.
This condition is satisfied only when $T^2=C_2^2=-1$ and $P^2=1$ for odd-$C_{2,6}$ pairing, which is the case of spin-orbit coupled systems we treat in the main text.
The low-energy Hamiltonian takes the form
\begin{align}
H=k_x\tau_z\sigma_x+k_y\tau_z\sigma_y
+\mu\tau_z+\Delta({\bf k})\tau_x\sigma_z,
\end{align}
where $\mu$ is the chemical potential, and $\Delta(-{\bf k})=-\Delta({\bf k})$. 
It is symmetric under $C_{2}=i\tau_z\sigma_z$, $T=i\sigma_yK$, and $P=\tau_y\sigma_yK$.
If we require $C_{6}=\tau_ze^{-i\frac{\pi}{6}\sigma_z}$ symmetry, an additional constraint $\Delta(R_{6}{\bf k})=-\Delta({\bf k})$ is imposed.

\subsubsection{Doublet and sextet of Majorana fermions}

Table~\ref{tab:charge} shows that point nodes at generic momenta can be protected when $(C_{2}T)^2=1$ and $(C_{2}P)^2=-1$, or $(C_{2}T)^2=-1$ and $(C_{2}P)^2=1$.
As we show below, rotation-protected Majorana fermions can appear in the former case, whereas they do not appear in the latter case.

Let us first consider the former case.
Since we consider odd-$C_{2,6}$ pairing with $C_{2}^2=-1$, we have $T^2=P^2=-1$.
We again introduce
\begin{align}
H_0=
k_x\sigma_x+k_y\tau_z\sigma_y.
\end{align}
Here, we take $T=i\sigma_yK$, $P=i\tau_yK$.
Then, two possible $C_{2}$ representations are $C_{2}=-i\sigma_z$ and $C_{2}=-i\tau_x\sigma_y$.
Those two representations are equivalent because they are related by a unitary transformation of the basis by $U=e^{-i\frac{\pi}{4}\tau_x\sigma_x}$.
We take $C_{2}=-i\sigma_z$.
Then, symmetry-allowed perturbations are
\begin{align}
\delta H
=m_1({\bf k})\tau_x\sigma_y+m_2({\bf k})\tau_y\sigma_z+m_3({\bf k})\tau_x+m_4({\bf k})\tau_z,
\end{align}
where $m_{1}(-{\bf k})=-m_{1}({\bf k})$, and $m_{2,3,4}(-{\bf k})=m_{2,3,4}({\bf k})$.
The $m_1$ term anticommutes with $H_0$, but it does not open the gap because it vanishes at ${\bf k}=0$.
The other terms are not mass terms, so they also do not open the gap.
Instead, $\delta H$ splits the fourfold degeneracy at ${\bf k}=0$ into a doublet of Majorana fermions.
If we require $C_{6}=\tau_ze^{-i\frac{\pi}{6}\tau_z\sigma_z}$ symmetry, $m_{1}(R_6{\bf k})=-m_{1}({\bf k})$, $m_{2,3}({\bf k})=0$, and $m_{4}(R_6{\bf k})=m_{4}({\bf k})$.
Then, a sextet of Majorana fermions appears at generic momenta by perturbations.

Next, we consider $(C_{2}T)^2=-1$ and $(C_{2}P)^2=1$.
Since $(C_{2}T)^2=-1$, $C_{2}T$ symmetry imposes Kramers degeneracy at each momentum, so the minimal Majorana Hamiltonian needs four bands and takes the form $H_{\rm min}=\rho_0\otimes (k_x\sigma_x+k_y\sigma_y)$, where $\rho_0$ is the $2\times 2$ identity matrix.
It has $C_{2}T=i\rho_y\sigma_xK$ and $C_{2}P=\rho_y\sigma_yK$ symmetries.
The Majorana fermion described by this Hamiltonian carries the U(1) winding number two.
Let us overlap two such Majorana fermions with opposite winding numbers to see whether a rotation-protected gapless spectrum can appear.
Its Hamiltonian has a $8\times 8$ form
\begin{align}
H_0=
k_x\sigma_x+k_y\tau_z\sigma_y.
\end{align}
There are four mass terms preserving $C_{2}T=i\rho_y\sigma_xK$ and $C_{2}P=i\rho_y\sigma_yK$ symmetries, which are
\begin{align}
\delta H_{\rm mass}
=(m_1\tau_x+m_2\tau_y\rho_x+m_3\tau_y\rho_y+m_4\tau_y\rho_z)\otimes \sigma_y,
\end{align}
where $m_{1,2,3,4}$ are constants.

We note that $C_{2}$ symmetry cannot forbid all mass terms.
If we require that $C_{2}$ symmetry disallows all mass terms, $C_{2}$ anticommutes with the mass terms as well as $H_0$.
Then, $C_{2}$ should be proportional to the chiral operator $S=\sigma_z$, but $C_{2}\propto S$ is incompatible with the odd-$C_{2}$ pairing condition.
Since $C_{2}$ anticommutes with $S=\sigma_z$ and $H_0$, it should be either
$C_{2}=-i\tau_y\sigma_y$, $C_{2}=-i\tau_x\rho_i\sigma_y$ for any $i=x,y,z$, or $C_{2}=-i\tau_x\sigma_y$.

Let us now show that $C_{6}$ symmetry also cannot forbid all mass terms.
When $C_{2}=-i\tau_y\sigma_y$, $C_{2}$ symmetry imposes $m_1=0$, and the mass term has the form $\delta H_{\rm mass}=(m_2\rho_x+m_3\rho_y+m_4\rho_z)C_{2}$.
In order for $C_{6}$ to eliminate $\delta H_{\rm mass}$, $C_{3}$ should act nontrivially on all Pauli matrices $\rho_{i=x,y,z}$ ($C_{2}$ acts trivially by definition).
However, it is impossible because a $C_{3}$ rotation leaves at least one dimension invariant in the three-dimensional $\rho_i$ space: the largest irreducible representation is two-dimensional, and the one-dimensional representation is trivial.
Thus, there exists a $C_{6}$-invariant mass term.
We can analyze the case with $C_{2}=-i\tau_x\rho_{i=x,y,z}\sigma_y$ in a similar way.
We choose $i=x$ for simplicity, but it is straightforward to consider other choices.
We have $m_2=0$ due to $C_{2}$ symmetry, and $\delta H_{\rm mass}=(m_1\rho_x+m_3\tau_z\rho_z+m_4\tau_z\rho_y)C_{2}$.
As in the previous case for $\rho_i$s, $\sigma_1\equiv \rho_x$, $\sigma_2\equiv \tau_z\rho_z$, and $\sigma_3\equiv \tau_z\rho_y$ satisfy the Lie algebra of Pauli matrices, $[\sigma_i,\sigma_j]=2i\epsilon_{ijk}\sigma_k$.
Again $C_{2}$ do not act on these Pauli matrices, and $C_{3}$ leaves at least one Pauli matrix invariant, so a mass term is allowed.
Finally, when $C_{2}=-i\tau_x\sigma_y$, the mass term $im_1C_{2}$ is invariant under $C_{6}$.
Thus, in all cases, the spectrum can be gapped in a way that respects $C_{6}$ symmetry and other internal symmetries $T$ and $S$.

\subsection{Odd-$C_{4}$ superconducting pairing}

As in the odd-$C_{6}$ pairing case, the only possibility is to have Majorana fermions at generic momenta with a trivial total winding number.
Odd-$C_{4}$ pairing is even-$C_{2}$, but a higher-spin spectrum cannot be protected because $C_{4}$ trivializes the 1D winding number in each $C_{2}$ eigensector.
Also, odd-$C_{4}$ symmetry forbids a nontrivial 3D winding number.
The anomalous surface states can thus appear as a quartet of Majorana fermions with zero total U(1) winding number or as a doubly charged Majorana Fermi line.

\subsubsection{Doubly charged Majorana Fermi lines}

A single doubly charged Majorana Fermi line can appear when $C_2^2=T^2=P^2=-1$ for odd-$C_{4}$ pairing.
This case is treated in the analysis of even-$C_{2}$ pairing above.
The only thing to check is whether it is compatible with $C_{4}$ symmetry.
One can see that the Hamiltonian Eq.~\eqref{eq:DCMFL_even} and symmetry representations $C_{2z}=i\tau_z\sigma_z$, $T=i\sigma_yK$, $P=i\tau_yK$ are compatible with $C_{4}=\tau_ze^{-i\frac{\pi}{4}\tau_z\sigma_z}$.

\subsubsection{Quartet of Majorana fermions}	

As we show in the main text, a quartet of Majorana fermions can be protected by $C_{4}$ symmetry in the same way as in rotation-protected topological crystalline insulators.
They are characterized by the $({\bb Z}_2)_{\rm M}$ invariant.

Such a quartet of Majorana fermions can be protected only in the spin-orbit coupled systems ($C_{4}^4=T^2=-1$ and $P^2=1$).
A point node at generic mometna can be protected when either $(C_{2}T)^2=1$ and $(C_{2}P)^2=-1$ or $(C_{2}T)^2=-1$ and $(C_{2}P)^2=1$ [Table.~\ref{tab:charge}].
In the former, the system is either spin-orbit coupled system or spin-SU(2)-symmetric system ($(C_{n})^n=T^2=1$ and $P^2=-1$).
Since a $d$-wave pairing in two-dimensional spin-SU(2)-symmetric systems produces four Majorana fermions in lattice systems, we know that there is no anomaly associated with the existence of four Majorana fermions.

$(C_{2}T)^2=-1$ and $(C_{2}P)^2=1$ with odd-$C_{4}$ pairing can be realized in exotic systems satisfying either $C_{4}^4=-1$, $T^2=1$ and $P^2=1$ or $C_{4}^4=1$, $T^2=-1$ and $P^2=1$. In these cases, $C_{2}T$ symmetry imposes Kramers degeneracy at every momenta on the $C_{4}$-invariant surface, and a stable Majorana point node has fourfold degenerate zero mode and carries an even number of winding number.
It is nontrivial to see whether a quartet of these Majorana fermions can be protected by crystalline symmetry.
Let us thus explicitly write down the low-energy effective Hamiltonian.

We first consider $C_{2z}T=i\sigma_yK$ and $C_{2z}P=K$ symmetries.
The most general form of the fourband Hamiltonian is $H({\bf k})=f_1({\bf k})\rho_y\sigma_x+f_2({\bf k})\rho_y\sigma_z$. We take $f_1=k_x$, $f_2=k_y$ to describe a Majorana fermion. It has winding number two.
Let us construct an eight-band Hamiltonian describing an overlap of two Majorana fermions with opposite winding numbers, which is
\begin{align}
H_0=k_x\rho_y\sigma_x+k_y\tau_z\rho_y\sigma_z.
\end{align}
Here, $\tau$ is the Pauli matrix for the flavor degrees of freedom.
$C_{2z}T$- and $C_{2z}P$-preserving mass terms are
\begin{align}
\delta H=m_1\tau_x\rho_y\sigma_z+m_2\tau_y\rho_z\sigma_x+m_3\tau_y\rho_x\sigma_x+m_4\tau_y\sigma_z.
\end{align}
Let us now see whether $C_{4z}$ that anticommutes with the particle-hole operator can exclude all mass terms.
A possible representation for $C_{2z}$ is a linear combination of elements in $\{1,\tau_z\}\otimes\{\rho_x, \rho_z, \rho_y\sigma_y\}$ when $C_{2z}^2=1$ and in $i\tau_y\otimes\{\rho_x, \rho_z, \rho_y\sigma_y\}\cup\{i\sigma_y\}$ when $C_{2z}^2=-1$, if we require that $C_{2z}H_0({\bf k})C_{2z}^{-1}=H(-{\bf k})$ and commutes with both $C_{2z}T$ and $C_{2z}P$.
Here, we can choose $\rho_x$ from $\{\rho_x, \rho_z, \rho_y\sigma_y\}$ by a transform of basis without affecting the form of $H_0$ and $\delta H$.
To see this, let us note that $\Gamma_1=\rho_x,\Gamma_2=\rho_z,\Gamma_3=\rho_y\sigma_x,\Gamma_4=\rho_y\sigma_y,\Gamma_5=\rho_y\sigma_z$ form five mutually anticommuting $4\times 4$ matrices.
Since $\{\rho_x, \rho_z, \rho_y\sigma_y\}=\{\Gamma_{1}, \Gamma_{2}, \Gamma_{4}\}$, we can rotate them fixing $\Gamma_3=\rho_y\sigma_x$ and $\Gamma_5=\rho_y\sigma_z$ (consider a $SO(2)$ rotation in a three-dimensional space with $\Gamma_1$, $\Gamma_2$, and $\Gamma_4$ as basis).
It leaves $H_0$ invariant, and the form of $\delta H$ is preserved after a redefinition of $m_2$ ,$m_3$, $m_4$.
Thus, we have without loss of generality that $C_{2z}=\rho_x$ or $\tau_z\rho_x$ when $C_{2z}^2=1$~\footnote{only $\theta=n\pi/2$ for an integer $n$ is allowed for $(\cos\theta+\sin\theta\tau_z)\rho_x$ if we require $C_{2z}^2=1$.} and $C_{2z}=i\tau_y\rho_x$ when $C_{2z}^2=-1$ (we do not consider $C_{2z}=i\sigma_y=iS$ because it is inconsisent with the assumption of odd-$C_{4z}$ pairing).

Let us see there are no protected surface states.
When $C_{2z}=\rho_x$, we have $m_1=m_2=0$.
Let us write the remaining mass term as $\delta H=\tau_y\rho_x\sigma_x(m_3-im_4\rho_x \sigma_y)$ and note that $\rho_x\sigma_y=C_{2z}S$ anticommutes with $C_{4z}$.
Since the $C_{4z}$ transformation of a mass term should be a mass term (because conjugating with $C_{4z}$ does not change the commutation relation), the only possibility is that either the $m_3$ term or the $m_4$ term is invariant under $C_{4z}$.
When $C_{2z}=\tau_z\rho_x$, $\delta H=\tau_x\rho_x\sigma_z(m_1+m_2C_{2z}S)$, and the same conclusion is derived.
When $C_{2z}=i\tau_y\rho_x$, there exists a unique choice
$C_{4z}=\tau_z\rho_y\sigma_ze^{-i\frac{\pi}{4}\tau_y\rho_x}$ that eliminates all the mass terms if we forget about symmetry of $H_0$.
This choice is, however, inconsistent with $C_{4z}$ symmetry of $H_0$.

\section{Spin-rotation-symmetric BdG Hamiltonian}
Let us investigate the effect of spin rotation symmetry on the BdG Hamiltonian following Ref.~\cite{schnyder2008classification}.
When a system has a spin-rotational symmetry around the $z$ axis, it is more effective to use the BdG Hamiltonian defined for each spin-$z$ eigensector separately.
If we assume a single pairing, the BdG Hamiltonian for the spin up sector has the following form
\begin{align}
H^{\uparrow\uparrow}_{\rm BdG}({\bf k})
=
\begin{pmatrix}
h_{\uparrow\uparrow}({\bf k}) & \Delta_{\uparrow\downarrow}({\bf k}) \\
(\Delta_{\uparrow\downarrow}({\bf k}))^{\dagger}  & -h_{\downarrow\downarrow}^t(-{\bf k})
\end{pmatrix}.
\end{align}
Here, the corresponding Nambu spinor is $\hat{\Psi}_{\uparrow}=(\hat{c}_{\uparrow\bf k},\hat{c}^{\dagger}_{\downarrow,-{\bf k}}$.
The spin-up BdG Hamiltonian belongs to class A because there is no nonspatial symmetry constraint.
If time reversal symmetry (under $\tilde{\cal T}=K$) is present, the symmetry constraint $[H^{\downarrow\downarrow}_{\rm BdG}({\bf k})]^*=H^{\uparrow\uparrow}_{\rm BdG}(-{\bf k})$ is identical to the chiral symmetry
\begin{align}
\tau_yH^{\uparrow\uparrow}_{\rm BdG}({\bf k})\tau_y^{-1}
=-H^{\uparrow\uparrow}_{\rm BdG}({\bf k}),
\end{align}
where $\tau_y$ is a Pauli matrix for the particle-hole indices.
Therefore, a spin-$z$-preserving superconductor with time reversal symmetry belongs to the AZ symmetry class AIII.

Let us now consider the full spin SU(2) rotation symmetry.
Since spin SU(2) rotation symmetry imposes that $h_{\uparrow\uparrow}({\bf k})=h_{\downarrow\downarrow}({\bf k})$, we can define the spinless BdG Hamiltonian as
\begin{align}
\tilde{H}_{\rm BdG}({\bf k})
\equiv
H^{\uparrow\uparrow}_{\rm BdG}({\bf k})
=
\begin{pmatrix}
\tilde{h}({\bf k}) & \tilde{\Delta}({\bf k}) \\
\tilde{\Delta}^{\dagger}({\bf k})  & -\tilde{h}^t(-{\bf k})
\end{pmatrix},
\end{align}
where we define $\tilde{h}({\bf k})\equiv h_{\uparrow\uparrow}({\bf k})=h_{\downarrow\downarrow}({\bf k})$ and $\tilde{\Delta}({\bf k})\equiv \Delta_{\uparrow\downarrow}({\bf k})$.
Here, compared to the case without spin rotation symmetry, we have a reversed sign for the condition that the pairing function satisfy: $\tilde{\Delta}({\bf k})=+^{\dagger}\tilde{\Delta}^t(-{\bf k})$.
Accordingly, the particle-hole operator for the spinless BdG Hamiltonian takes the form
\begin{align}
\tilde{P}=
\begin{pmatrix}
0&1\\
1&0
\end{pmatrix}
K,
\end{align}
which satisfy $\tilde{P}^2=-1$.
Thus, a spin-SU(2)-symmetric BdG Hamiltonian is in class C.
When time reversal symmetry is present, the combination of time reversal and a spin $\pi$-rotation defines effective spinless time reversal symmetry satisfying $\tilde{T}^2=+1$, so the BdG Hamiltonian belongs to class CI.

Up to now, we assume spin-singlet pairing.
In spin-polarized normal metals, however, triplet pairing should occur, so the relevant BdG Hamiltonian takes the form
\begin{align}
H^{\uparrow\uparrow}_{\rm BdG}({\bf k})
=
\begin{pmatrix}
h_{\uparrow\uparrow}({\bf k}) & \Delta_{\uparrow\uparrow}({\bf k}) \\
(\Delta_{\uparrow\uparrow}({\bf k}))^{\dagger}  & -h_{\uparrow\uparrow}^t(-{\bf k})
\end{pmatrix},
\end{align}
where we suppose that spins are polarized along the $\uparrow$ direction.
It has particle-hole symmetry under
\begin{align}
\tilde{P}=
\begin{pmatrix}
0&1\\
1&0
\end{pmatrix}
K,
\end{align}
where $\tilde{P}^2=1$. While the spin polarization breaks time reversal symmetry and spin rotation symmetry individually, there remains a symmetry under the combination of time reversal and a spin rotation around the axis perpendicular to the spin-ordering axis.
Since this effective time reversal $\tilde{T}$ satisfies $\tilde{T}^2=+1$, the BdG Hamiltonian belongs to the class BDI.

\section{ A model of spin-polarized $C_{2}$-protected topological superconductor}

\begin{figure}[b]
\includegraphics[width=8.5cm]{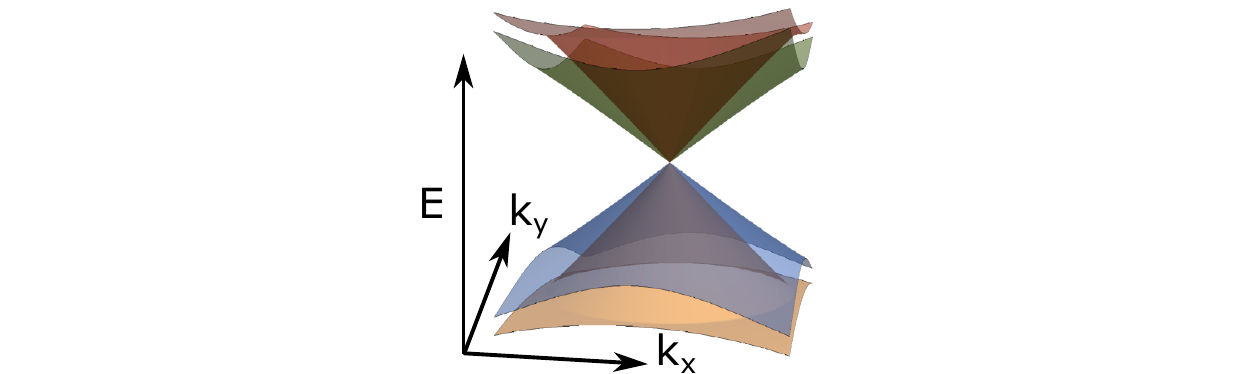}
\caption{
Surface spectrum of a spin-polarized $C_{2}$-protected topological superconductor.
We use the model in Eq.~\eqref{eq:model-spin-pol} with $\mu=-0.2$, $m_1=m_2=0.2$, and $\Delta=1$.
The system has $20$ unit cells along the $z$ direction and periodic boundary condition along the $xy$ direction.
}
\label{fig:numerical-supp}
\end{figure}

Here we construct a model of spin-polarized $C_{2}$-protected topological superconductor hosting a spin-3/2 fermion on the surface.
This system belongs to the class $C_{2}^2=T^2=P^2=1$.

We consider the following normal-state Hamiltonian.
\begin{align}
\label{eq:model-spin-pol}
h
=\mu+f_1 \rho_y\sigma_x+f_2\sigma_y + f_3\sigma_z
+ m_1\rho_z+m_2\rho_z\sigma_z,
\end{align}
where $f_1=\sin k_x$, $f_2=\sin k_y$, and $f_3=2-\cos k_x-\cos k_y-\cos k_z$.
It is symmetric under ${\cal C}_{2z}={\cal I}=\sigma_z$ and ${\cal T}=K$: ${\cal C}_{2z}h({\bf k})({\cal C}_{2z})^{-1}=h(-k_x,-k_y,k_z)$, ${\cal I}h({\bf k}){\cal I}^{-1}=h(-{\bf k})$, ${\cal T}h({\bf k}){\cal T}^{-1}=h(-{\bf k})$.
This Hamiltonian describes two Dirac points carrying nontrivial ${\bb Z}_2$ monopole charges at $k_z=\pm \pi/2$ when $m_1=m_2=0$, and the Dirac poins become nodal lines with a finite size when $m_1,m_2\ne 0$~\cite{fang2015topological,ahn2018band}.
We open the bulk superconducting gap fully by introducing even-$C_{2}$ pairing $\Delta({\bf k})=-i\Delta\sin k_z$.
When $|\mu|$, $|m_1|$, and $|m_2|$ are small, this pairing introduces winding number $+1$ ($-1$) for each band within the $C_{2}=+1$ ($-1$) sector along the $(k_x,k_y,k_z)=(0,0,k_z)$ line.
Note that our normal state Hamiltonian and pairing function are both diagonal along this line. The winding number of the sector with $\rho_z=s_{\rho}=\pm 1$ and $\sigma_z=s_{\sigma}=\pm 1$ is given by the phase winding number of $E_{\rho\sigma}({\bf k})+\Delta_{\rho\sigma}({\bf k})=\mu+s_{\rho} m_1+s_{\rho}s_{\sigma}m_2-s_{\sigma}\cos k_z-i\Delta \sin k_z$, which is just $s_{\sigma}$ when $|\mu|$, $|m_1|$, $|m_2|$ are much smaller than one.
So, we have $w_{\rm rot}=2$, while $w_{\rm rot}=0$ for other $C_{2}$-invariant lines because no Fermi surface is there.
We show the numerically calculated spectrum for $\mu=-0.2$, $m_1=m_2=0.2$, and $\Delta=1$ in Fig.~\ref{fig:numerical-supp}, where one can observe the anomalous surface state with a fourfold degenerate crossing point at $(k_x,k_y)=(0,0)$.

Let us comment on the previous work on two-dimensional spin-3/2 fermions in the literature.
About a decade ago, four spin-3/2 fermions called {\it birefringent fermions} (described by a variant of Eq.~\eqref{eq:model-spin-pol}) were shown to appear in the Brillouin zone of a 2D square optical lattice~\cite{kennett2011birefringent}.
This seems like indicating that a non-superconducting (and non-anomalous) realization of spin-3/2 fermions is possible in two dimensions.
However, those birefringent fermions are not strictly protected by the symmetry of ordinary complex fermionic (i.e., non-Majorana) systems.
Our analysis shows that their topological protection requires sublattice symmetry ($=$chiral symmetry), which is present only in nearest-neighbor hopping models, as well as $C_{2}$ and $T$ symmetries.
Therefore, to our knowledge, the Majorana fermion proposed here is the first as well as anomalous realization of a symmetry-protected birefringent fermion in two dimensions.

\section{Berry phase on time-reversal-invariant loops}
\label{sec:Berry}

Here we show that the Berry phase on a time-reversal-invariant loop is always trivial in spinless systems with time reversal symmetry.
From this, it follows that a single doubly charged Majorana Fermi line cannot appear in spinless systems, because it originates from the nontrivial Berry phase in the normal state.

\subsection{Constraint from time reversal symmetry}

Before we study the Berry phase around a time-reversal-invariant loop, let us first summarize the definition and some useful properties of the abelian Berry connection $A({\bf k})$ for the occupied states (i.e., $E<0$ states in the Bogoliubov-de Gennes formalism).
The abelian Berry connection is defined by
\begin{align}
A({\bf k})=\sum_{n\in {\rm occ}}\braket{u_{n{\bf k}}|i\nabla_{\bf k}|u_{n{\bf k}}}.
\end{align}
When the matrix element of time reversal operator at ${\bf k}$ is given by
\begin{align}
B_{mn}({\bf k})=\braket{u_{m-{\bf k}}|T|u_{n{\bf k}}},
\end{align}
which satisfies
\begin{align}
\label{eq:sewing-symmetry}
B(-{\bf k})=(-1)^{s_T} B^t({\bf k})
\end{align}
when $T^2=(-1)^{s_T}=\pm 1$,
the Berry connetion satisfies the following constraint
\begin{align}
A({\bf k})=A(-{\bf k})-i\nabla_{\bf k}\log \det B({\bf k}).
\end{align}

Now we suppose that the band gap is open on a time-reversal-invariant loop $l=l_0+Tl_0$, where $l_0$ is an arc whose starting- and end-points are nonzero ${\bf k}_0$ and $-{\bf k}_0$.
Then, the Berry phase $\Phi_l$ on the loop $l$ is
\begin{align}
\Phi_l
&=\oint_l d{\bf k}\cdot A({\bf k})\notag\\
&=\int_{l_0} d{\bf k}\cdot A({\bf k})
+\int_{Tl_0} d{\bf k}\cdot A({\bf k})\notag\\
&=\int_{l_0} d{\bf k}\cdot [A(-{\bf k})-i\nabla_{\bf k}\log \det B({\bf k})]
+\int_{Tl_0} d{\bf k}\cdot A({\bf k})\notag\\
&=-i\int^{-{\bf k}_0}_{{\bf k}_0} d{\bf k}\cdot \nabla_{\bf k}\log \det B({\bf k})\notag\\
&=-i\log \left[\frac{\det B(-{\bf k}_0)}{\det B({\bf k}_0)}\right]\notag\\
&=-i\log \left[(-1)^{N_{\rm occ}s_T}\right]\notag\\
&=\pi N_{\rm occ}s_T \mod 2\pi,
\end{align}
where we use in the fourth line that $\oint_{l_0} d{\bf k}\cdot A(-{\bf k})=-\oint_{Tl_0} d{(T\bf k)}\cdot A(T{\bf k})=-\oint_{Tl_0} d{\bf k}\cdot A({\bf k})$, $N_{\rm occ}$ is the number of the occupied states on the loop and use Eq.~\eqref{eq:sewing-symmetry} in the sixth line.

Our derivation shows that in spinless systems where $T^2=1$, the Berry phase around a time-reversal-invariant loop is always zero (mod $2\pi$).
Thus, band crossing carrying Berry phase $\pi$ (in particular, a 2D Dirac point) cannot appear at a time-reversal-invariant momentum (TRIM) in spinless systems.

Alternatively, the same conclusion can be drawn for a 2D Dirac point by investigating the Hamiltonian.
We consider the most general form of an effective two-level Hamiltonian around a TRIM (${\bf k}=0$).
\begin{align}
H({\bf k})=\mu({\bf k})+f_1({\bf k})\sigma_x+f_2({\bf k})\sigma_y+f_3({\bf k})\sigma_z.
\end{align}
If we take a basis where $T=K$ without loss of generality, we find
\begin{align}
\mu(k_x,k_y)&=\mu(-k_x,-k_y)\notag\\
f_{1,3}(k_x,k_y)&=f_{1,3}(-k_x,-k_y)\notag\\
f_2(k_x,k_y)&=-f_2(-k_x,-k_y).
\end{align}
Since only $f_2$ is an odd function of ${\bf k}$, while a Dirac point at a TRIM requires two independent components of ${\bf f}=(f_1,f_2,f_3)$ to be odd in ${\bf k}$, we cannot achieve a Dirac point even after considering any other symmetry.

\subsection{Relation to the 1D charge of a Majorana Fermi line}

We can relate the Berry phase of a Dirac fermion in the normal state to the 1D topological charge of a Majorana Fermi line as follows.
Let us recall that a Majorana Fermi line can be realized on the surface of a topological superconductor when $(C_{2}T)^2=(C_{2}P)^2=1$ and $(C_{2}T)S=S(C_{2}T)$ on the surface, where $S$ is the Hermitian chiral operator [See Sec.~\ref{sec:classification} and Table.~\ref{tab:symmetry}].
If we choose a basis in which $S={\rm diag}[1_{N\times N},-1_{N\times N}]$ and $C_{2}T=K$ where $1_{N\times N}$ denotes the $N\times N$ identity matrix, 
the surface BdG Hamiltonian takes the following off-diagonal form due to $SH_{\rm BdG}({\bf k})+H_{\rm BdG}({\bf k})S=0$:
\begin{align}
\label{eq:BDI-Ham}
H_{\rm BdG}({\bf k})
=
\begin{pmatrix}
0& O({\bf k})\\
O^{\dagger}({\bf k}) & 0
\end{pmatrix},
\end{align}
where $O({\bf k})$ is real-valued because of $C_{2}T=K$ symmetry, 
so that it can be continuously deformed to an element of the orthogonal group $O(N)$ without closing the band gap.
The 0D and 1D topological charges of Majorana Fermi lines are given by the zeroth and first homotopy classes of $O({\bf k})$, which are $\pi_0[O(N)]=\pi_1[O(N)]={\mathbb Z}_2$ when $N\ge 3$ while $\pi_0[O(N)]=\pi_1[O(N)]={\mathbb Z}$ when $N=2$~\cite{bzdusek2017robust,kawakami2019topological}. 
Considering that $O({\bf k})$ corresponds to the normal-state Hamiltonian $h({\bf k})$ when $\Delta=0$, one can see that the topological charges of the Majorana Fermi line (mod 2) originate from the corresponding topological charges of the Fermi lines in the normal state in the weak pairing case.

\section{Symmetry of the winding numbers}
\label{sec:winding}

Here, we study the symmetry properties of the 1D and 3D winding numbers under crystalline and time reversal symmetries.

Let $G$ be a point group symmetry operator.
Then, the 3D winding number in $G$-symmetric systems satisfies
\begin{align}
w_{\rm 3D}
&=\frac{1}{48\pi^2}\int_{\rm BZ} d^3k
{\rm Tr}\left[S(H^{-1}\nabla_{\bf k} H)^3\right]\notag\\
&=\frac{1}{48\pi^2}\int_{\rm BZ} d^3k
{\rm Tr}\left[(GSG^{-1})(H^{-1}(G{\bf k})\nabla_{\bf k}H(G{\bf k}))^3\right]\notag\\
&=\frac{1}{48\pi^2}\int_{\rm BZ} d^3k\epsilon^{ijk}\left(\det{G}\right)
{\rm Tr}\left[(GSG^{-1})(H^{-1}\nabla H)^3\right]_{G\bf k}\notag\\
&=\frac{\left(\det{G}\right)}{48\pi^2}\int_{\rm BZ} d^3k
{\rm Tr}\left[(GSG^{-1})(H^{-1}\nabla_{\bf k}H)^3\right]\notag\\
&=\pm\left(\det{G}\right) w_{\rm 3D}.
\end{align}
where $(H^{-1}\nabla_{\bf k}H)^3=\epsilon^{ijk}H^{-1}\partial_{k_i}HH^{-1}\partial_{k_j}HH^{-1}\partial_{k_k}H$, the Hamiltonian $H$ is $G$-symmetric, and the sign in the last line indicates the $G$-parity of the pairing function $U_G\Delta({\bf k})U_G^{-1}=\Delta(R_G{\bf k})$, where $R_G{\bf k}$ is the natural transformation of ${\bf k}$ under $G$.
It shows that $w_{\rm 3D}=0$ for odd-$C_{n}$ pairing, even-$M$, or even-$I$ pairing.

This is consistent with the constraint on the sum of winding numbers carried by MFs on $G$-invariant surfaces for $G=C_{n}$ or $M$.
Let $l$ be a $G$-invariant (invariant up to orientation reversal) loop in a 2D Brillouin zone.
The winding number $w_l$ around the loop then satisfies
\begin{align}
w_l
&=\frac{i}{4\pi}\oint_l d{\bf k}\cdot{\rm Tr}\left[SH^{-1}\nabla_{\bf k}H\right]\notag\\
&=\frac{i}{4\pi}\oint_l d{\bf k}\cdot{\rm Tr}\left[SG^{-1}H^{-1}(G{\bf k})G\nabla_{\bf k}G^{-1}H(G{\bf k})G\right]\notag\\
&=\frac{i}{4\pi}\oint_l d{\bf k}\cdot{\rm Tr}\left[(GSG^{-1})H^{-1}(G{\bf k})\nabla_{\bf k}H(G{\bf k})\right]\notag\\
&=\frac{i}{4\pi}\oint_{G\cdot l} d(G{\bf k})\cdot{\rm Tr}\left[(GSG^{-1})H^{-1}(G{\bf k})\nabla_{G\bf k}H(G{\bf k})\right]\notag\\
&=\frac{i}{4\pi}\left(\det{G}\right)\oint_l d{\bf k}\cdot{\rm Tr}\left[(GSG^{-1})H^{-1}\nabla_{\bf k}H\right]\notag\\
&=\pm \left(\det{G}\right)w_l,
\end{align}
where we use that $\oint_{G\cdot l}=\left(\det{G}\right)\oint_l$ in the fifth line.
Since $w_l=0$ for any $G$-invariant loop in the case of odd-$C_{n}$ or even-$M$ pairing, the total winding number in the surface Brillouin zone should be zero.

Let us also consider the 1D winding number $w_{\rm 1D}$ defined on a noncontractible loop, which is a 1D Brillouin zone.
\begin{align}
w_{\rm 1D}
&=\frac{i}{4\pi}\int_{\rm 1D\;BZ} dk{\rm Tr}\left[SH^{-1}\nabla_kH\right].
\end{align}
We have
\begin{align}
\label{eq:1Dwinding-G}
w_{\rm 1D}
&=\pm s_Gw_{\rm 1D},
\end{align}
where the sign in fron of $s_G$ captures the $G$-parity of the pairing function, and $s_G=\pm 1$ is defined by $Gk=s_Gk$.

Similarly, one can show that for time reversal symmetry,
\begin{align}
w_{\rm 3D}
&=\frac{1}{48\pi^2}\int_{\rm BZ} d^3k{\rm Tr}\left[S(H^{-1}\nabla_{\bf k}H)^3\right]\notag\\
&=-\frac{1}{48\pi^2}\int_{\rm BZ} d^3k
{\rm Tr}\left[(T^{-1}ST)(H^{-1}\nabla_{\bf k}H)^3\right].
\end{align}
Note that the requirement $S^2=1$, needed for the above expression of the winding number, is satisfied when we define $S$ as
\begin{align}
S
&=i^{s_T+s_C}TC,
\end{align}
with the convention $TC=CT$, where $T^2=(-1)^{s_T}$ and $C^2=(-1)^{s_C}$.
Therefore,
\begin{align}
T^{-1}ST
&=(-1)^{s_T+s_C}S.
\end{align}
Thus, we have
\begin{align}
w_{\rm 3D}
&=-(-1)^{s_T+s_C}w_{\rm 3D}.
\end{align}
This shows that the winding number is trivial in the classes BDI and CII where $(-1)^{s_T+s_C}=+1$.
One can also derive
\begin{align}
w_{l}
&=-(-1)^{s_T+s_C}w_{l}.
\end{align}

In contrast, we have
\begin{align}
w_{\rm 1D}
&=+(-1)^{s_T+s_C}w_{\rm 1D},
\end{align}
such that the winding number in the 1D Brillouin zone is trivial in the classes DIII and CI.
More generally, if we consider a $C_{n}$-invariant line, the winding number $w^{\lambda}_{\rm 1D}$ for each eigensector of $C_{n}$ rotational symmetry with eigenvalue $\lambda$ satisfies
\begin{align}
\label{eq:1Dwinding-eig-T}
w^{\lambda}_{\rm 1D}
&=+(-1)^{s_T+s_C}w^{\lambda^*}_{\rm 1D}.
\end{align}

\section{Majorana Kramers pairs in spin-orbit coupled 2D systems}
Here, we show the emergence of Majorana Kramers pairs on the boundary of superconductors obtained by odd-$C_{n=2,4,6}$ pairing in doped 2D ${\mathbb Z}_2$ topological insulators.
To investigate the boundary states, let us begin with a low-energy effective Hamiltonian of the ${\mathbb Z}_2$ topological insulator.
\begin{align}
h({\bf k})
&=-\mu+M\rho_z+k_x\rho_x\sigma_z+k_y\rho_y,
\end{align}
where $\rho_{i=x,y,z}$ and $\sigma_{i=x,y,z}$ are the Pauli matrices for the orbital and spin degrees of freedom.
It is symmetric under
\begin{align}
{\cal T}
=i\sigma_yK,\quad
{\cal C}_{n}
=e^{-i\frac{\pi}{n}\rho_z\sigma_z}.
\end{align}
The corresponding Bogoliubov-de Gennes (BdG) Hamiltonian has the form of the doubled ${\mathbb Z}_2$ topological superconductor.
\begin{align}
H_{\rm BdG}({\bf k})
&=M\tau_z\rho_z+k_x\tau_z\rho_x\sigma_z+k_y\tau_z\rho_y\notag\\
&=M\Gamma_1+k_x\Gamma_2+k_y\Gamma_3,
\end{align}
where $\tau_{i=0,x,y,z}$ are Pauli matrices for the Nambu space, and we introduce $8\times 8$ mutually anticommuting Gamma matrices
\begin{align}
\Gamma_1
&=\tau_z\rho_z
&(+,-,+,+),\notag\\
\Gamma_2
&=\tau_z\rho_x\sigma_z
&(-,+,-,-),\notag\\
\Gamma_3
&=\tau_z\rho_y
&(-,+,-,-),\notag\\
\Gamma_4
&=\tau_z\rho_x\sigma_x
&(-,+,-,-),\notag\\
\Gamma_5
&=\tau_z\rho_x\sigma_y
&(-,+,-,-),\notag\\
\Gamma_6
&=\tau_x
&(+,-,+,-),\notag\\
\Gamma_7
&=\tau_y
&(+,+,+,-),
\end{align}
where the four signs show their commutation $(+)$ or anticommutation $(-)$ relations with
\begin{align}
T
&=i\sigma_yK,\notag\\
C
&=\tau_y\sigma_yK,\notag\\
C^{\rm even}_{2z}
&=-i\rho_z\sigma_z,\notag\\
C^{\rm odd}_{2z}
&=-i\tau_z\rho_z\sigma_z,
\end{align}
in order.
One can see that a superconducting gap $\Delta \Gamma_6$ is allowed for even-$C_{2}$ pairing, which shows that the bulk topology is trivial.
However, $C_{4}$ symmetry forbids the bulk mass term because $C^{\rm odd}_{4z}
=\tau_ze^{-i\frac{\pi}{n}\rho_z\sigma_z}$ anticommutes with it.
Therefore, the mass term is prohibited in the translation-invariant bulk for any odd-$C_{n=2,4,6}$ pairing.
However, $\Delta \Gamma_6$ can describe the pairing gap on the boundary.
We can see this by allowing the position dependence of the pairing in real space~\cite{khalaf2018higher,geier2018second}.
\begin{align}
H({\bf k},{\bf r})=M({\bf r})\Gamma_1+k_x\Gamma_2+k_y\Gamma_3+\Delta({\bf r})\Gamma_6.
\end{align}
Here, $M({\bf r})$ is the bulk mass term which approaches to a constant value deep in the bulk and to zero on the boundary, and $\Delta({\bf r})$ is the surface pairing function that is nonzero only near the boundary, where translation symmetries are broken.
Since $\Delta(R_{n}{\bf r})=-\Delta({\bf r})$ for odd-$C_{n}$ pairing, the pairing gap has to close at $n$ (mod $2n$) points on the boundary.
These are the locations where Majorana zero modes appear (often called Majorana corner modes~\cite{khalaf2018higher,geier2018second} because they usually appear at corners of the system~\cite{hwang2019fragile}).
The zero modes form a Kramers pair at each site because of time reversal symmetry.
This shows that odd-$C_{n}$ pairing on the helical edge states of a ${\mathbb Z}_2$ topological insulator induces Majorana Kramers pairs on the boundary.

\section{Topological charges for a $C_n$-multiplet of spin-half fermions}
\label{sec:Qn}

Here we explain the topological invariant protecting the multiplet of Majorana fermions [a quartet of Majorana fermions is shown in Fig.~\ref{fig:quartet}(a) as an example].
While such a configuration appears in $C_{4}$-symmetric systems in ordinary superconductors, a similar configuration with two and six surface nodal points can appear in $C_{2}$- and $C_{6}$-symmetric topological crystalline insulators or spin-orbit coupled $C_{2}$- and $C_{6}$-symmetric superconductors with exotic particle-hole symmetry where $P^2=-1$ [See Table.~\ref{tab:symmetry}].
With this in mind, we consider general $C_{n=2,4,6}$-symmetric systems.
Since particle-hole symmetry does not play an important role here, the results here can be applied to the surface states of topological crystalline insulators also.

\subsection{General formulation}

Before we study the topological invariant around a $C_{n}$-invariant loop, let us first see how $C_{n}$ symmetry constrains the abelian Berry connection $A({\bf k})$ for the occupied states.
When the matrix element of $C_{n}$ operator is given by
\begin{align}
D_{mn}({\bf k})=\braket{u_{mR_{n}{\bf k}}|C_{n}|u_{n{\bf k}}},
\end{align}
the Berry connetion satisfies the following constraint
\begin{align}
A({\bf k})=R_{n}^{-1}\cdot A(R_{n}{\bf k})+i\nabla_{\bf k}\log \det D({\bf k}).
\end{align}

\begin{figure}[t!]
\includegraphics[width=8.5cm]{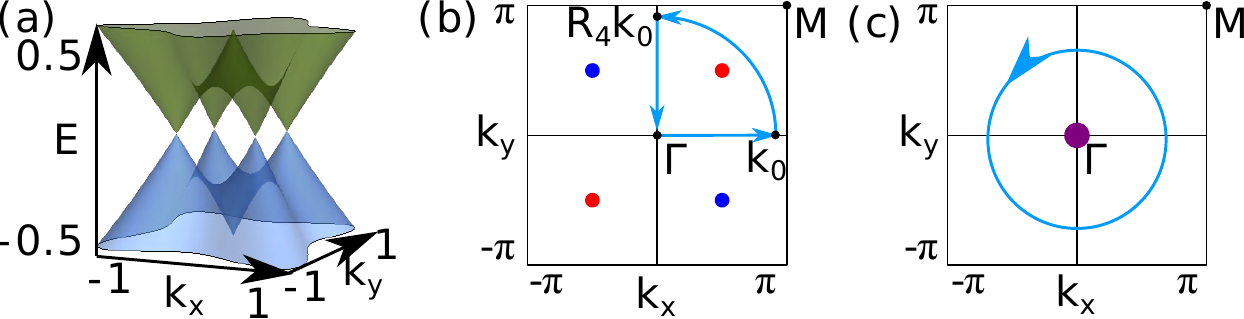}
\caption{
A quartet of Majorana fermions (QMF) for odd-$C_{4z}$ pairing.
(a) Energy spectrum of the BdG Hamiltonian $H_{\rm BdG}(k_x,k_y)=k_x\tau_z\sigma_x+k_y\tau_z\sigma_y+\delta H_{\Delta}$, where $\delta H_{\Delta}
=\Delta_1\tau_x+\Delta_2\tau_x\sigma_x+\Delta_3\tau_x\sigma_y$ with $\mu=0.5$, $\Delta_1=0$, $\Delta_2=0.5k_x$, $\Delta_3=-0.5k_y$
on a $C_{4z}$ invariant surface.
(b) Generic configuration of Majorana fermions in the surface Brillouin zone.
Red (blue) dots indicate Majorana fermions with the winding number $+1$ ($-1$).
The Berry phase calculated over the closed trajectory connected by blue arrows characterizes the surface anomaly.
(c) Four Majorana fermions merged at the $\Gamma$ point.
Their global stability is guaranteed by the topological charge $Q_4$ in Eq.~(\ref{eq:Qn}) defined along the circle (light blue) enclosing the $\Gamma$ point.
}
\label{fig:quartet}
\end{figure}

In spin-orbit coupled systems with both $C_{n=2,4,6}$ and time reversal symmetries, we can define a ${\mathbb Z}_2$ topological invariant $Q_n$ over a $1/n$ segment of a $C_{n}$-invariant loop.
We consider the Berry phase around a closed loop connecting $\Gamma=(0,0)$, ${\bf k}_0$, and $R_{n}{\bf k}_0$ for a nonzero ${\bf k}_0$ [Fig.~\ref{fig:quartet}(b)].
\begin{align}
&\Phi_{1/n}^{{\bf k}_0}\notag\\
&=\int^{{\bf k}_0}_{\Gamma}d{\bf k}\cdot{\bf A}({\bf k})
+\int^{R_{n}{\bf k}_0}_{{\bf k}_0}d{\bf k}\cdot{\bf A}({\bf k})
+\int^{\Gamma}_{R_{n}{\bf k}_0}d{\bf k}\cdot{\bf A}({\bf k})\notag\\
&=\int^{{\bf k}_0}_{\Gamma}d{\bf k}\cdot{\bf A}({\bf k})
+\int^{R_{n}{\bf k}_0}_{{\bf k}_0}d{\bf k}\cdot{\bf A}({\bf k})
+\int^{\Gamma}_{{\bf k}_0}d{\bf k}\cdot R_{n}^{-1}{\bf A}(R_{n}{\bf k})\notag\\
&=\int^{{\bf k}_0}_{\Gamma}d{\bf k}i\nabla_{\bf k}\log\det D({\bf k})
+\int^{R_{n}{\bf k}_0}_{{\bf k}_0}d{\bf k}\cdot{\bf A}({\bf k})\notag\\
&=\pi Q_n^{{\bf k}_0}-i\nabla_{\bf k}\log\det D(\Gamma),
\end{align}
where we define $Q_n^{{\bf k}_0}$ on a line connecting ${\bf k}_0$ and $R_{n}{\bf k}_0$:
\begin{align}
\label{eq:Qn}
Q_n^{{\bf k}_0}
&=\frac{i}{\pi}\log\det D({\bf k}_0)+\frac{1}{\pi}\int^{R_{n}{\bf k}_0}_{{\bf k}_0} d{\bf k}\cdot A({\bf k}).
\end{align}
This quantity does not depend on the choice of the initial point ${\bf k}_0$: $Q_{{\bf k}'_0}-Q_n^{{\bf k}_0}=0$ on a given loop.
Let us thus omit the subscript ${\bf k}_0$ for $Q_n^{{\bf k}_0}$.
Also, $Q_n$ is gauge invariant under $\ket{u_{n\bf k}}\rightarrow U_{mn}\ket{u_{m\bf k}}$, which transforms $D$ and $A$ by $D({\bf k})\rightarrow U^{-1}(R_{n}{\bf k})D({\bf k})U({\bf k})$ and $A({\bf k})\rightarrow A({\bf k})+i\nabla_{\bf k}\log \det U({\bf k})$.

We can see that $Q_n$ is quantized in the presence of time reversal symmetry by taking a real gauge where $C_{2}T\ket{u_{n\bf k}}=\ket{u_{n\bf k}}$.
In a real gauge, $A({\bf k})=0$ and $D({\bf k})$ belongs to an orthogonal group such that $\det D=\pm 1$.
From this, it follows that $Q_n=\pm 1$ in a real gauge.
Moreover, since $Q_n$ is gauge invariant,
\begin{align}
Q_n=\pm 1
\end{align}
holds in any gauge.

As is clear from the definition, $Q_{n}$ is closely related to $\Phi^{{\bf k}_0}_{1/n}$, and they are equivalent for the purpose of studying the local stability of gapless points within the $1/n$ sector of the Brillouin zone.
In fact, they are identical when the spectrum is gapped at $\Gamma$ because $\det D(\Gamma)=1$ due to Kramers degeneracy.
However, when we study the global stability of gapless surface states protected by $C_{n}$ symmetry, we need to use the invariant $Q_n$ rather than the quantized Berry phase $\Phi^{{\bf k}_0}_{1/n}$.
It is because $\Phi^{{\bf k}_0}_{1/n}$ cannot be defined in the case where all the gap closing points gather at the $\Gamma$ point [Fig.~\ref{fig:quartet}(c)].
To see whether such a critical gapless point at $\Gamma$ can be gapped or not, we have to look at $Q_n$, which is well-defined on a fully gapped loop surrounding $\Gamma$.

This ${\mathbb Z}_2$ invariant $Q_n$ is the topological invariant carried by the gapless surface states of $C_{n=2,4,6}$-protected topological insulators proposed by Fang and Fu~\cite{fang2019new}, while the authors did not figure out this topological invariant and relied on the low-energy effective Hamiltonian to study the protection of the gapless states.
Let us take the real gauge such that $Q_n=(i/\pi)\log\det D({\bf k}_0)$.
If we contract the loop where $Q_{n}$ is defined to a TRIM ${\bf k}_{\rm TRIM}$, we always have $Q_{n}=0$ in lattice systems because of the Kramers degeneracy, imposing $\det D({\bf k}_{\rm TRIM})=1$.
Thus, $Q_n=1$ reveals the rotation anomaly on the surface of topological crystalline insulators or superconductors.
We remark that, while gapped spin-orbit coupled lattice systems always have $Q_{n}=0$ because $\det D=1$ by time reversal symmetry, $Q_{n}$ can change by a band inversion in spinless systems.
So, there is no surface anomaly due to nontrivial $Q_{n}$ in spinless systems.

Our invariant $Q_{n}$ is equivalent to the Wilson line invariant defined to capture fragile topology in $C_{6}$-symmetric systems in Ref.~\cite{bradlyn2019disconnected,bouhon2019wilson}.
Let us define the Wilson line operator for the occupied states on the line connecting ${\bf k}_2$ and ${\bf k}_1$ by
\begin{align}
W_{{\bf k}_2\leftarrow {\bf k}_1}
&=\lim_{{\bm \delta}\rightarrow 0}
P_{{\bf k}_2}P_{{\bf k}'-{\bm \delta}}...P_{{\bf k}+{\bm \delta}}P_{{\bf k}_1},
\end{align}
where
\begin{align}
P_{\bf k}
=\sum_{n\in{\rm occ}}\ket{u_{n{\bf k}}}\bra{u_{n\bf k}}
\end{align}
is the projection to the occupied states at momentum ${\bf k}$.
Since the projection satisfies $P_{R_{n}\bf k}=C_{n}\sum_{n\in{\rm occ}}\ket{u_{n{\bf k}}}\bra{u_{n\bf k}}C_{n}^{-1}$ in $C_{n}$-symmetric systems, the Wilson line has the following property.
\begin{align}
W_{R_{n}{\bf k}_2\leftarrow R_{n}{\bf k}_1}
&=
C_{n}W_{{\bf k}_2\leftarrow {\bf k}_1}C_{n}^{-1}
\end{align}
Thus, the Wilson loop operator over a $C_{n}$-invariant loop with initial point ${\bf k}_0$ is given by
\begin{align}
W_{{\bf k}_0\leftarrow {\bf k}_0}
&=W_{{\bf k}_0\leftarrow C_{n}^{n-1}{\bf k}_0}\hdots W_{R_{n}{\bf k}_0\leftarrow {\bf k}_0}\notag\\
&=C_{n}^{n-1}W_{R_{n}{\bf k}_0\leftarrow {\bf k}_0}C_{n}^{-(n-1)}\hdots W_{R_{n}{\bf k}_0\leftarrow {\bf k}_0}\notag\\
&=(-1)^{s_T}
(C_{n}^{-1}W_{R_{n}{\bf k}_0\leftarrow {\bf k}_0})^n,
\end{align}
where we use $C_{n}^{n-1}=(-1)^{s_T}C_{n}^{-1}$.
Then, if we define the {\it $C_{n}$-Wilson line} by
\begin{align}
W^{C_{n}}_{mn}
\equiv \braket{u_{m{\bf k}_0}|C_{n}W_{R_{n}{\bf k}_0\leftarrow {\bf k}_0}|u_{n{\bf k}_0}},
\end{align}
it is independent of ${\bf k}_0$ and gauge invariant up to similarity transformations, such that its spectrum is gauge invariant.
This $C_{n}$-Wilson line is related to $Q_n$ by
\begin{align}
e^{iQ_n}
=\det W^{C_{n}}.
\end{align}

\subsection{Comment on the superconducting case}

As we state in the main text, the global stability of the Majorana fermions still has a ${\mathbb Z}_2$ character although the presence of $S$ symmetry promotes the ${\mathbb Z}_2$-valued Berry phase of twofold degenerate Majorana fermions to the integer-valued winding number.
To understand this, let us suppose that there are eight Majorana fermions on the surface BZ so that a quadrant of the BZ has two MFs with the same chirality. 
Since the operation of $C_{4}$ changes the sign of the winding number for odd-$C_{4}$ pairing, MFs in the adjacent quadrants should have an opposite sign of winding numbers.
Then, by continuously shifting the positions of Majorana fermions while keeping $T$, $S$, and $C_{4}$ symmetries, we can annihilate MFs pairwise at the borders between neighboring quadrants of the BZ. 
On the other hand, when the number of Majorana fermions on the surface is four, such a pair-annihilation process is impossible, and thus QMF is stable.

\subsection{Explicit calculation for a spin-half fermion}

$Q_n$ can be exactly calculated analytically in the Dirac (or Majorana) point limit.
Let us first begin with a twofold degenerate Dirac point at the $\Gamma$ point described by
\begin{align}
h_{m}
&=
k
\begin{pmatrix}
0&e^{-im\theta}\\
e^{im\theta}&0
\end{pmatrix}.
\end{align}
Af half filling, one state is occupied at nonzero ${\bf k}$, and it takes the form
\begin{align}
\ket{u^{m}_{\rm occ}}
=
\frac{1}{\sqrt{2}}
\begin{pmatrix}
1\\
-e^{im\theta}
\end{pmatrix}.
\end{align}
$h_{m}$ is symmetric under the $n$-fold rotation
\begin{align}
C^{s,m}_{n}
&=(-1)^{s}\exp\left(-i\frac{m\pi}{n}\sigma_z\right).
\end{align}
Namely, $C^{s,m}_{n}h_{m}(k,\theta)(C^{s,m}_{n})^{-1}=h_{m}(k,\theta+\pi/n)$.
The matrix element of the rotation operator is given by
\begin{align}
D^{s,m}_n({\bf k}_0)
&=\braket{u^{m}_{R_{n}{\bf k}}|C^{s,m}_{n}|u^{m}_{\bf k}}
=(-1)^{s}e^{-\pi im/n}.
\end{align}
Thus,
\begin{align}
\frac{i}{\pi}\log\det D^{s,m}_{n}({\bf k}_0)
=
-s+\frac{m}{n}.
\end{align}
On the other hand, the Berry connection is given by
\begin{align}
A_{\theta}({\bf k})
&=
\braket{u^{\pm}_{\rm occ}|i\partial_{\theta}|u^{\pm}_{\rm occ}}
=-\frac{m}{2},\notag\\
A_{k}({\bf k})
&=
\braket{u^{\pm}_{\rm occ}|i\partial_{k}|u^{\pm}_{\rm occ}}
=0,
\end{align}
such that
\begin{align}
\frac{1}{\pi}\int^{R_{n}{\bf k}_0}_{{\bf k}_0}d{\bf k}\cdot A({\bf k})
=-\frac{m}{n}.
\end{align}
We have
\begin{align}
Q_n
=
s
\mod 2.
\end{align}
Thus, in the case of fourfold Dirac point formed by two identical twofold Dirac points,
\begin{align}
h_{4\times 4}=h_{m}\oplus h_{l},
\end{align}
the rotation-protected topological charge is given by
\begin{align}
\label{eq:X-4Dirac}
Q_n=s+s'\mod 2,
\end{align}
where $s$ and $s'$ are defined for the rotation operator acting on $h_m$ and $h_l$, respectively.
It is consistent with the analysis based on the low-energy effective Hamiltonian in Ref.~\cite{fang2019new}.

\subsection{Lattice model calculation for a rotation-protected topological insulator}

\begin{figure}[t!]
\includegraphics[width=8.5cm]{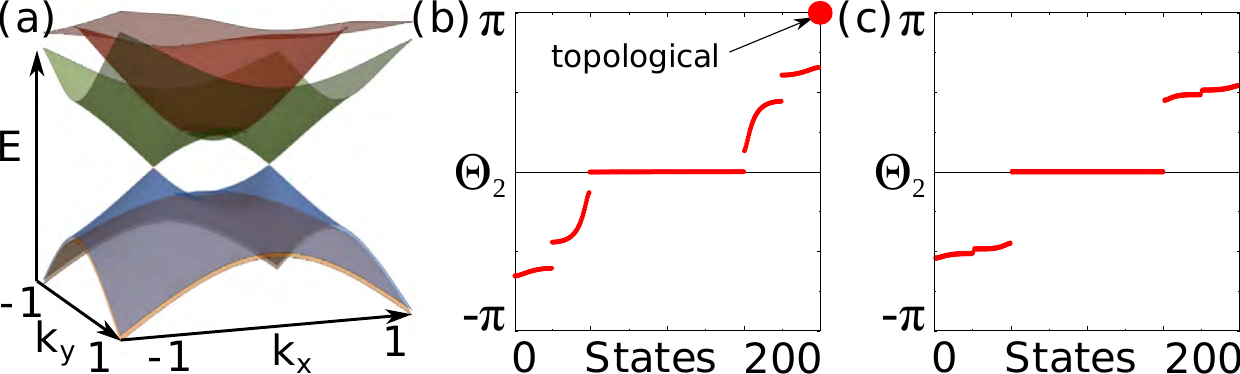}
\caption{
$C_{2}$-protected topological crystalline insulator.
The system has a finite number --- 20 in (a) and 50 in (b) and (c) --- of unit cells along $z$ with open boundaries, while it is periodic along $x$ and $y$.
(a) Low-energy surface spectrum near $(k_x,k_y)=(0,0)$.
(b,c) $C_{2}$-Wilson line spectrum for the whole system with 200 occupied bands.
They are calculated over an $\pi$-arc of the circle $k_x^2+k_y^2=1$ in (b), enclosing two Dirac points, and $k_x^2+k_y^2=0.3^2$ in (c), enclosing no Dirac points.
Anomalous gapless states on each of the top and bottom surfaces give the eigenvalue $\Theta_2=\pi$ when the circle encloses them, such that the whole system has two eigenstates with $\Theta_2=\pi$.
Thus, $Q_{2}=e^{i\Theta_2}=-1$ on the circle $k_x^2+k_y^2=1$ for each of the top and bottom surfaces.
}
\label{fig:numerical}
\end{figure}

In the above, we show that $C_{n}$-protected surface states of the Fang-Fu type~\cite{fang2019new} (a $C_{n}$-multiplet of Dirac/Majorana fermions) are characterized by the topological invariant $Q_n$.
To confirm this further, let us consider a lattice model of $C_{2}$-protected topological crystalline insulator.
We begin with a strong topological insulator protected by time reversal symmetry.
\begin{align}
H^{\rm STI}_{\pm}
&=
\sin k_x \rho_x\sigma_x
\pm \sin k_y \rho_x\sigma_y
+\sin k_z \rho_x\sigma_z\notag\\
&+(2-\cos k_x-\cos k_y-\cos k_z) \rho_z,
\end{align}
where $\rho_{i=x,y,z}$ and $\sigma_{i=x,y,z}$ are Pauli matrices for the orbital and spin degrees of freedom.
This system has symmetries under ${\cal T}=i\sigma_yK$, ${\cal C}_{2z}=-i\sigma_z$, and ${\cal I}=\rho_z$.
It has one Dirac cone on each surface whose chirality depending on the sign $\pm$.
From this we construct the Hamiltonian of a $C_{2}$-protected topological insulator as
\begin{align}
H
&=H^{\rm STI}_{+}\oplus H^{\rm STI}_{-}.
\end{align}
Let us preseve ${\cal C}_{2z}=(-i\sigma_z)\oplus (-i\sigma_z)$ symmetry of this system by adding $C_{2z}$- and $T$-preserving perturbations (we also keep $P$ symmetry for convenience of analysis below)
\begin{align}
\delta H=m_1\tau_x\rho_z+m_2\tau_z\rho_z+m_3\tau_x+m_4\tau_z
\end{align}
to the Hamiltonian $H$, where $\tau_{i=x,y,z}$ are Pauli matrices for the chirality degrees of freedom $+$ and $-$.
Then, we have two Dirac cones on each $C_{2}$-preserving surface as shown in Fig.~\ref{fig:numerical}(a).

Since our choice of $C_{2z}$ has the same representations for the two surface Dirac cones with opposite chirality, the rotation-protected topological charge $Q_n$ is nontrivial as shown in Sec.~\ref{sec:Qn}.
To see this numerically, we take the open boundary condition along $z$ and the periodic boundary condition along $x$ and $y$.
Then we calculate the spectrum $\Theta_2$ of the $C_{2}$-Wilson line around an $\pi$-arc on the circle $|{\bf k}|=1$ from $\theta=0$ to $\theta=\pi$ for the whole system.
Figure~\ref{fig:numerical}(b) shows that there appears two eigenstates of the $C_{2}$-Wilson line with $\theta_2=\pi$.
Since the bulk is gapped, this nontrivial value should be attributed to the gapless surface states.
One is from the top surface, and the other is from the bottom surface because of inversion symmetry.
Thus, we conclude that $Q_2=1 \mod 2$ for each of the top and bottom surface states.

\section{Classification of pairing functions in lattice models}

\begin{table}[b!]
\begin{tabular}{cccccccc}
$\Delta$	&Matrix					& Irrep			&$C_4$ 	&$M_z$	&$M_x$	&$M_y$ 	&$M_{x+y}$\\
\hhline{========}
$\Delta_{1a}$		&$\rho_0\sigma_0$		&$A_{1g}$		&$+$		&$+$		&$+$		&$+$		&$+$\\
$\Delta_{1b}$		&$\rho_z\sigma_0$		&$A_{1g}$		&$+$		&$+$		&$+$		&$+$		&$+$\\
$\Delta_{2}$		&$\rho_y\sigma_y$		&$B_{1u}$		&$-$		&$-$		&$-$		&$-$		&$-$\\
$\Delta_{3}$		&$\rho_y\sigma_x$		&$B_{2u}$		&$-$		&$-$		&$+$		&$+$		&$+$\\
$\Delta_{4a}$		&$\rho_x\sigma_0$		&$E_{u}$		&$-\Delta_{4b}$		&$+$		&$+$		&$-$		&$-\Delta_{4b}$\\
$\Delta_{4b}$		&$\rho_y\sigma_z$		&$E_{u}$		&$\Delta_{4a}$		&$+$		&$-$		&$+$		&$-\Delta_{4a}$\\
\end{tabular}
\caption{
Time-reversal-preserving constant pairing functions of the Dirac semimetal $h_1$.
}
\label{tab:pairing}
\end{table}

\begin{table}[b!]
\begin{tabular}{cccccccc}
$\Delta$	&Matrix					& Irrep			&$C_4$ 	&$M_z$	&$M_x$	&$M_y$ 	&$M_{x+y}$\\
\hhline{========}
$\Delta_{1a'}$		&$\rho_0\sigma_z$		&$A_{2g}$		&$+$		&$+$		&$-$		&$-$		&$-$\\
$\Delta_{1b'}$		&$\rho_z\sigma_z$		&$A_{2g}$		&$+$		&$+$		&$-$		&$-$		&$-$\\
$\Delta_{2'}$		&$\rho_x\sigma_y$		&$B_{1u}$		&$-$		&$-$		&$-$		&$-$		&$-$\\
$\Delta_{3'}$		&$\rho_x\sigma_x$		&$B_{2u}$		&$-$		&$-$		&$+$		&$+$		&$+$\\
$\Delta_{4a'}$		&$\rho_0\sigma_y$		&$E_{g}$		&$-\Delta_{4b'}$		&$-$		&$-$		&$+$		&$-\Delta_{4b'}$\\
$\Delta_{4b'}$		&$\rho_z\sigma_x$		&$E_{g}$		&$\Delta_{4a'}$		&$-$		&$+$		&$-$		&$-\Delta_{4a'}$\\
$\Delta_{5a'}$		&$\rho_z\sigma_y$		&$E_{g}$		&$-\Delta_{5b'}$		&$-$		&$-$		&$+$		&$-\Delta_{5b'}$\\
$\Delta_{5b'}$		&$\rho_0\sigma_x$		&$E_{g}$		&$\Delta_{5a'}$		&$-$		&$+$		&$-$		&$-\Delta_{5a'}$\\
$\Delta_{6a'}$		&$\rho_y\sigma_0$		&$E_{u}$		&$-\Delta_{6b'}$		&$+$		&$+$		&$-$		&$-\Delta_{6b'}$\\
$\Delta_{6b'}$		&$-\rho_x\sigma_z$		&$E_{u}$		&$\Delta_{6a'}$		&$+$		&$-$		&$+$		&$-\Delta_{6a'}$\\
\end{tabular}
\caption{
Time-reversal-breaking constant pairing functions of the Dirac semimetal $h_1$.
These constant pairing functions should be multiplied by a ${\bf k}$-odd function to preserve time reversal symmetry.
}
\label{tab:pairing2}
\end{table}

\subsection{Dirac-semimetal-based model}

We consider two lattice models in the main text.
One is a Dirac semimetal with gap-opening terms.
It is
\begin{align}
\label{eq:modelH1}
h_1
=-\mu
+f_1\rho_z+f_2\rho_x\sigma_z+f_3\rho_y+f_4\rho_x\sigma_x+f_5\rho_x\sigma_y,
\end{align}
where $\rho_{i=x,y,z}$ and $\sigma_{i=x,y,z}$ are Pauli matrices for orbital and spin degrees of freedom, respectively, and $f_1=4-2(\cos k_x+\cos k_y)-\cos k_z$, $f_2=\sin k_x$, $f_3=-\sin k_y$, $f_4=3\sin k_z(\cos k_y-\cos k_x)+m_0 \sin k_z$, and $f_5=-\sin k_z\sin k_x\sin k_y+m_1 \sin k_z$.
$m_0$ and $m_1$ are gap-opening terms that break $C_{4z}$, $M_x$, $M_{x+y}$, and $M_y$ symmetries.
When $m_0=m_1=0$, $h_1$ has $T$ and $D_{4h}$ symmetries under
\begin{align}
{\cal T}
&=i\sigma_yK,\notag\\
{\cal M}_x
&=i\sigma_x,\notag\\
{\cal M}_y
&=i\rho_z\sigma_y,\notag\\
{\cal M}_{x+y}
&=\frac{1}{\sqrt{2}}(\sigma_x-\rho_z\sigma_y),\notag\\
{\cal M}_z
&=i\sigma_z,\notag\\
{\cal C}_{4z}
&=e^{-i(\pi/4)(2\rho_0-\rho_z)s_z}.
\end{align}
We classify superconducting pairing functions based on their transformation properties under $D_{4h}$ symmetry group operations in Tables.~\ref{tab:pairing} and~\ref{tab:pairing2}.

As we state in the main text, $\Delta({\bf k})=\Delta_{1a'}\sin k_z=\Delta_e\sin k_z\sigma_z$ is the unique even-$C_{2z}$ pairing function that leads to a rotation-protected topological superconductor.
Let us explain why it is so.
They are $\Delta_{1a}$, $\Delta_{1b}$, $\Delta_{2}$, $\Delta_{3}$ in Table~\ref{tab:pairing}, and $\Delta_{1a'}\sin k_z$, $\Delta_{1b'}\sin k_z$, $\Delta_{2'}\sin k_z$, $\Delta_{3'}\sin k_z$ from Table~\ref{tab:pairing2} if we consider the lowest Fourier expansions in $k_z$ (we neglect $\sin k_x$ and $\sin k_y$ terms because they vanish at $C_{2z}$-invariant lines).
$\Delta_{2}$, $\Delta_{3}$, $\Delta_{2'}\sin k_z$ and $\Delta_{3'}\sin k_z$ are odd-$C_{4z}$ pairing, so they are also incompatible with higher-spin Majorana fermions.
$\Delta_{1a}$ and $\Delta_{1b}$ are even-parity pairing, so they are incompatible with higher-spin Majorana fermions as we discuss in the main text.
The only remaining are $\Delta_{1a'}\sin k_z$ and $\Delta_{1b'}\sin k_z$, which has the same symmetry properties
However, the latter leads to a topologically trivial phase at the $C_{4}$-invariant line, and in fact, it does not open the bulk gap.
Let us thus investigate the $\Delta_{1a'}$ more closely to see whether it really leads to $w_{+}=-w_-=2$ along the ${\bf k}=(0,0,k_z)$ line, such that a HSMF appears at the surface momentum ${\bf k}_s=(0,0)$.
The 1D BdG Hamiltonian along the ${\bf k}=(0,0,k_z)$ line is $H_{\rm BdG}=-\mu\tau_z-\cos k_z\rho_{z}\tau_z+\Delta_{1a'} \sin k_z \sigma_{z}\tau_x$, where $\tau_{i=x,y,z}$ are Pauli matrices for particle-hole indices, and $m_0=m_1=0$ is assumed for simplicity.
Since $H_{\rm BdG}$ is diagonal in the spin and orbital indices, it can be labelled as $H^{s_{\rho},s_{\sigma}}_{\rm BdG}$ where $s_{\rho}$ and $s_{\sigma}$ indicate the eigenvalues of $\rho_z$ and $\sigma_z$, respectively.
Then, one can show that each $2\times 2$ matrix $H^{s_{\rho},s_{\sigma}}_{\rm BdG}$ has the winding number $w_{s_{\rho},s_{\sigma}}=-s_{\rho}s_{\sigma}$ for $S=\tau_y$, which gives $w_{+}=-w_{-}=2$ for $C_{2z}=-i\rho_z\sigma_z$.

\begin{table}[b!]
\begin{tabular}{cccccccc}
$\Delta$	&Matrix					& Irrep			&$C_4$ 	&$M_z$	&$M_x$	&$M_y$ 	&$M_{x+y}$\\
\hhline{========}
$\Delta_{1a}$		&$\rho_0\sigma_0$		&$A_{1g}$		&$+$		&$+$		&$+$		&$+$		&$+$\\
$\Delta_{1b}$		&$\rho_z\sigma_0$		&$A_{1g}$		&$+$		&$+$		&$+$		&$+$		&$+$\\
$\Delta_{2}$		&$\rho_y\sigma_z$		&$A_{1u}$		&$+$		&$-$		&$-$		&$-$		&$+$\\
$\Delta_{3}$		&$\rho_x\sigma_0$		&$A_{2u}$		&$+$		&$-$		&$+$		&$+$		&$-$\\
$\Delta_{4a}$		&$\rho_y\sigma_y$		&$E_{u}$		&$-\Delta_{4a}$		&$+$		&$-$		&$+$		&$\Delta_{4b}$\\
$\Delta_{4b}$		&$\rho_y\sigma_x$		&$E_{u}$		&$\Delta_{4b}$		&$+$		&$+$		&$-$		&$\Delta_{4a}$\\
\end{tabular}
\caption{
Time-reversal-preserving constant pairing functions of the gapped nodal line semimetal model described by $h_2$.
}
\label{tab:pairing3}
\end{table}

\subsection{Nodal-line-semimetal-based model}

The other model describes a nodal line semimetal gapped by spin-orbit coupling.
The Hamiltonian is
\begin{align}
\label{eq:modelH2}
h_{2}=&-\mu +\sin k_z\rho_y + (M-\sum_{i=x,y,z}\cos k_i)\rho_z
\nonumber\\
&+\lambda_{{\rm SO}}(\sin k_x\rho_x\sigma_y-\sin k_y\rho_x\sigma_x),
\end{align}
which is symmetric under 
\begin{align}
{\cal T}
&=i\sigma_yK,\notag\\
{\cal M}_x
&=i\sigma_x,\notag\\
{\cal M}_y
&=i\sigma_y,\notag\\
{\cal M}_{x+y}
&=\frac{i}{\sqrt{2}}(\sigma_x+\sigma_y),\notag\\
{\cal M}_z
&=i\rho_z\sigma_z,\notag\\
{\cal C}_{4z}
&=e^{-i\frac{\pi}{4}\sigma_z}.
\end{align}
Superconducting pairing functions are classified according to transformation properties under $D_{4h}$ symmetry group operations in Tables~\ref{tab:pairing3}, and~\ref{tab:pairing4}.

Let us note that not every odd-$C_{4z}$ pairing gives a quartet of Majorana fermions on the $C_{4z}$-invariant surface.
We show in the main text that a d-wave pairing $\Delta_2(\cos k_y-\cos k_x)+\Delta_3\sin k_x\sin k_y$ gives the quartet of Majorana fermions.
In contrast, other odd-$C_{4}$ pairing $k_x\Delta_{4a'}+k_y\Delta_{4b'}$ or $k_x\Delta_{5a'}+k_y\Delta_{5b'}$ opens the full gap on the $C_{4z}$-invariant surface.
We also note that the above $d$-wave pairing does not open the gap on the side surfaces.
This is because $\Delta_d({\bf k})$, which is odd under $C_{4z}$, is also odd under $M_z$, so that there appear gapless Majorana surface states protected by $M_z$ symmetry on the side surfaces.
Therefore, to open the side surface gap and observe the hinge states, we should add an additional odd-$C_4$ pairing term such as $\delta\Delta({\bf k})\propto \sin k_x\rho_x\sigma_y+\sin k_y\rho_x\sigma_x$ which is even under $M_z$.

\begin{table}[hb!]
\begin{tabular}{cccccccc}
$\Delta$	&Matrix					& Irrep			&$C_4$ 	&$M_z$	&$M_x$	&$M_y$ 	&$M_{x+y}$\\
\hhline{========}
$\Delta_{1a'}$		&$\rho_0\sigma_z$		&$A_{2g}$		&$+$		&$+$		&$-$		&$-$		&$-$\\
$\Delta_{1b'}$		&$\rho_z\sigma_z$		&$A_{2g}$		&$+$		&$+$		&$-$		&$-$		&$-$\\
$\Delta_{2'}$		&$\rho_x\sigma_z$		&$A_{1u}$		&$+$		&$-$		&$-$		&$-$		&$-$\\
$\Delta_{3'}$		&$\rho_y\sigma_0$		&$A_{2u}$		&$+$		&$-$		&$+$		&$+$		&$+$\\
$\Delta_{4a'}$		&$\rho_0\sigma_y$		&$E_{g}$		&$-\Delta_{4b'}$		&$-$		&$-$		&$+$		&$\Delta_{4b'}$\\
$\Delta_{4b'}$		&$\rho_0\sigma_x$		&$E_{g}$		&$\Delta_{4a'}$		&$-$		&$+$		&$-$		&$\Delta_{4a'}$\\
$\Delta_{5a'}$		&$\rho_z\sigma_y$		&$E_{g}$		&$-\Delta_{5b'}$		&$-$		&$-$		&$+$		&$\Delta_{5b'}$\\
$\Delta_{5b'}$		&$\rho_z\sigma_x$		&$E_{g}$		&$\Delta_{5a'}$		&$-$		&$+$		&$-$		&$\Delta_{5a'}$\\
$\Delta_{6a'}$		&$\rho_x\sigma_y$		&$E_u$		&$-\Delta_{6b'}$		&$+$		&$-$		&$+$		&$\Delta_{6b'}$\\
$\Delta_{6b'}$		&$\rho_x\sigma_x$		&$E_u$		&$\Delta_{6a'}$		&$+$		&$+$		&$-$		&$\Delta_{6a'}$
\end{tabular}
\caption{
Time-reversal-breaking constant pairing functions of the gapped nodal line semimetal model described by $h_2$.
These constant pairing functions should be multiplied by a ${\bf k}$-odd function to preserve time reversal symmetry.
}
\label{tab:pairing4}
\end{table}

\end{document}